\def\C{\overline{\tt c}}
\def\S{\overline{\tt s}}
\def\EE{\overline{\tt ee}}
\def\EC{\overline{\tt ec}}
\def\ES{\overline{\tt es}}
\def\CE{\overline{\tt ce}}
\def\CC{\overline{\tt cc}}
\def\CS{\overline{\tt cs}}
\def\SE{\overline{\tt se}}
\def\SC{\overline{\tt sc}}
\def\SSS{\overline{\tt ss}}
\def\X{{\rm X}}
\def\Y{{\rm Y}}

\def\tr{\mathop{\rm tr}}

\documentclass[10pt,prb,aps,twocolumn,showpacs,superscriptaddress]{revtex4}
\usepackage{epsf}
\usepackage{graphicx,color}
\usepackage{amssymb}
\usepackage{subfigure}
\usepackage[hypertex]{hyperref}
\def\urlprefix#1#2{\hskip0pt plus0.01fil\discretionary{}{}{}%
  {\url{#2}}}
\begin{document}

\renewcommand{\topfraction}{0.95}        
\renewcommand{\bottomfraction}{0.8}     
\setcounter{topnumber}{2}
\setcounter{bottomnumber}{2}
\setcounter{totalnumber}{4}     
\setcounter{dbltopnumber}{2}    
\renewcommand{\dbltopfraction}{0.95}     
\renewcommand{\textfraction}{0.07}      
\renewcommand{\floatpagefraction}{0.9}  
\renewcommand{\dblfloatpagefraction}{0.9}       

\title{Second-order shaped pulses for solid-state quantum computation}

\author{Leonid P. Pryadko} 
\affiliation{Dept. of Physics, University of California, Riverside, CA 92521}
\author{Pinaki Sengupta} 
\affiliation{T-CNLS and  MPA-NHMFL, Los Alamos National Laboratory, Los
  Alamos, NM 87545} 

\date{\today}

\begin{abstract}
We present the construction and detailed analysis of highly-optimized
self-refocusing pulse shapes for several rotation angles.  We characterize the
constructed pulses by the coefficients appearing in the Magnus expansion up to
second order.  This allows a semi-analytical analysis of the performance of
the constructed shapes in sequences and composite pulses by computing the
corresponding leading-order error operators.  Higher orders can be analyzed
with the numerical technique suggested by us previously.  We illustrate the
technique by analyzing several composite pulses designed to protect against
pulse amplitude errors, and on decoupling sequences for potentially long
chains of qubits with on-site and nearest-neighbor couplings.
\end{abstract}

\maketitle

\section{Introduction} 
The implementation of quantum algorithms using NMR on molecules in
liquid\cite{chuang-2001}, solid\cite{muller-2003}, and liquid
crystals\cite{liq-xtal-nmr} has demonstrated {\em in principle\/} that
pulse-based control methods can be useful for quantum information
processing\cite{vandersypen-2004} (QIP).  The technique has long been
a staple in NMR spectroscopy, where complex molecules like proteins
are analyzed with the help of long sequences of precisely designed
radiofrequency pulses\cite{slichter-book}.  Related techniques for
coherent manipulation of many-body quantum systems have emerged as an
important tool in many areas of science and technology.

A useful quantum computer should contain hundreds, if not thousands of qubits.
The only hope of scaling to such system sizes is with the help of multiple
levels of quantum error correction (QEC).  For this to work, the benefits due
to each additional level of encoding should outweigh the corresponding
overhead of additional errors.  This leads to various threshold
theorems\cite{Knill-error-bound,Fowler-QEC-2004}, estimating the maximum error
rate for which such concatenated encoding can be beneficial.  The
corresponding thresholds are rather stringent, meaning that for scalability
one needs very accurate elementary gates.

Even for relatively small $n$-body systems, the number of states
scales exponentially with $n$, and the accuracy required for QIP is
high.  As demonstrated in several recent experiments in NMR QIP,
required accuracy can be reached with the help of {\em
  strongly-modulating pulses\/}, where entire single- or few-qubit
gates are designed numerically for a given
molecule\cite{price-havel-cory-2000,boulant-2003} [also see
  \onlinecite{sanders-1999-complex-instruct,%
    hodgkinson-2000,niskanen-2003,steffen-2007}].  While the technique indeed
offers unprecedentedly accurate, fast gates (which also helps to avoid
relaxation), it obviously cannot be generalized to larger systems.

In contrast, the traditional pulse and sequence design rely on the
Magnus (cumulant) expansion\cite{slichter-book}.  The expansion is
done around the evolution in the applied controlling fields, while the
chemical shifts\cite{warren-herm, bauer-gauss,geen-freeman-burp}
(resonant-frequency offsets) and inter-qubit couplings are treated
perturbatively.  The main advantage of the Magnus expansion is its
locality.  Namely, when local qubit couplings are dominant, the
control fields accurate to a given order can be designed by analyzing
relatively small clusters.  The results remain exact independent of
the system size, or even in the limit of infinite system.  One can
thus characterize pulse-based method for designing control fields as
{\em scalable\/} to large system sizes.
\cite{goswami-review}.

A scheme to systematically construct high-order self-refocusing
pulses and pulse sequences was developed 
by the authors in Ref.~\onlinecite{sengupta-pryadko-ref-2005}.
Specifically, we constructed ``soft''
NMR-style\cite{warren-herm,freeman-shaped-pulse} second-order self-refocusing
inversion ($\pi$) pulses and several high-order sequences based on such pulses for
refocusing qubits arranged in spin chains with on-site chemical shifts and XXZ
nearest-neighbor coupling.  The main technical advance which enabled the
calculations\cite{sengupta-pryadko-ref-2005} was the  efficient
numerical algorithm for computing high-order terms of the Magnus expansion.
The algorithm is based on the usual time-dependent perturbation theory; the
direct computation of multiple integrals entering higher-order cumulants would
be totally impossible.

In this paper we present highly-optimized self-refocusing pulse shapes for
rotation angles $\phi_0$ other than 180$^\circ$.  For such pulses, we extend
the results of Ref.~\onlinecite{pryadko-quiroz-2007} and construct the
analytical expansion of the evolution operator for an arbitrary coupled qubit.
While the expansion is more complicated than that for the inversion pulses
with $\phi_0=180^\circ$, to second order, it is still characterized by only
three coefficients, two of which we suppress by pulse shaping.  This allows us
to compute the error operators associated with a given control sequence
semi-analytically, by evaluating the leading order terms in the corresponding
products of the evolution operators.  We illustrate the technique on several
newly-constructed decoupling sequences for a chain of qubits with on-site and
nearest-neighbor couplings, as well as with the composite pulses protecting
against amplitude errors.
\section{Problem setup}
\subsection{Dynamical Decoupling}

In principle, the simplest type of control pulses consist of short,
intense bursts of coherent, resonant electromagnetic radiation,
popularly known as ``hard'' or ``bang-bang'' pulses.  In this limit,
for the duration $\tau$ of the pulse one can totally ignore all other
couplings of the qubit(s).  Then, a pulse sequence can be viewed as a
series of free evolution intervals [unitary evolution operators
$U_f(t)=\exp(-i t H_S)$, where $H_S$ is the system Hamiltonian]
intercalated with pulse operators.  For example, with single-qubit
control Hamiltonian,
\begin{equation}
  H_C={1\over2}\sum_a V^\mu_a(t)\sigma^\mu_a, 
  \label{eq:control-1}
\end{equation}
where $\sigma^\mu_a$, $\mu=x,y,z$ are the Pauli matrices for $a$-th
qubit and $V^\mu_a(t)$ are the corresponding control
fields (or, more precisely, the envelopes of the
  control fields applied at the resonant frequency of the
  corresponding qubit), the pulse operator is the product of those for
individual qubits, $P=\prod_a P_a(\phi_0^{(a)},\hat{\bf n}_a) $,
\begin{equation}
  \label{eq:pulse-general}
  P_a(\phi_0,{\bf n})
  =\cos{\phi_0\over2}-i \hat {\bf
  n}\cdot \vec\sigma_a\sin {\phi_0\over2}.
\end{equation}
Here $\hat{\bf n}$ is the unit vector that determines the spin rotation
axis and  $\phi_0$ is the corresponding angle, 
$$\phi_0\, \hat{\bf
  n}^\mu=\int_0^\tau dt\, V^\mu(t).$$

The corresponding pulse algebra is straightforward.  For example, for a
single spin with the chemical shift Hamiltonian,
\begin{equation}
  H_S={\Delta\over 2} \,\sigma^z,
  \label{eq:chemical-shift}
\end{equation}
the sequence of two equally-spaced inversion pulses ($\phi_0=\pm\pi$)
in $x$-direction is equivalent to a unitary,
\begin{eqnarray}
  \lefteqn{P(\pi,\hat{\bf x}) \,
   U_f(t)\,
   P(-\pi,\hat{\bf x}) \,
   U_f(t)}& & \nonumber \\ 
   & &=(-i\sigma^x)\, \exp\Bigl(-i{t\Delta \over2}  \sigma^z\Bigr)
   (i\sigma^x)\, \exp\Bigl(-i{t\Delta \over2}  \sigma^z\Bigr)
   \nonumber\\
   & & = \exp\Bigl(+i{t\Delta \over2}  \sigma^z\Bigr)
   \exp\Bigl(-i{t\Delta \over2}  \sigma^z\Bigr)=\openone , 
   \label{eq:spin-echo-unitary}
\end{eqnarray}
which is, of course, the formal identity behind the well-known spin-echo
sequence\cite{hahn-1950}.

In reality, the pulse duration $\tau$ is always finite.  Thus, during the
pulse application, the rotation actually happens around the axis determined by
the sum of the system and control Hamiltonians; one gets finite-pulse-duration
errors.  Generically, such errors scale {\em linearly\/} with pulse duration.
These errors are especially dangerous in systems where one cannot reduce the
pulse duration indefinitely, e.g., because of the need for spectral addressing
in homonuclear NMR, or in order to avoid exciting levels outside the qubit
state in Josephson phase qubits\cite{steffen-2003}.  Yet, even in systems with
optical qubit coupling where $\tau$ could be in the sub-picosecond range, the
finite-pulse-duration errors can be significant if one targets the accuracy
necessary for achieving the scalability
thresholds\cite{chen-piermarocchi-sham-2001}.

The finite-pulse-duration errors can be significantly reduced by
suppressing the leading-order error operators.  The errors would
scale with a higher power of the pulse duration, which makes them much
more manageable.  This can be achieved either by designing special
sequences (which typically doubles or quadruples the number of
pulses), or by using specially designed self-refocusing pulse
shapes\cite{warren-herm}.  Typically, the latter strategy is more
efficient\cite{sengupta-pryadko-ref-2005}; besides, the
self-refocusing pulses can often be used as drop-in replacements for
corresponding
$\delta$-pulses\cite{sengupta-pryadko-ref-2005,pryadko-quiroz-2007}.

\subsection{Model}

We consider the following simplified Hamiltonian
\begin{equation}
  \label{eq:ham-total}
   H(t)= H_{\rm C}(t)+ H_{\rm S}+ H_{V}(t)+ H_{\sigma},
\end{equation}
with the first (main) term  due to individual control fields, 
\begin{equation}
  \label{eq:ham-control}
   H_{\rm C}(t)={1\over2}\sum_n \bigl[ V^x_n(t)\,
  \sigma^x_n+V^y_n(t)\, \sigma^y_n\bigr] , 
\end{equation}
where $\sigma_n^\mu$, $\mu=x,y,z$, are the usual Pauli matrices
for the $n$-th qubit (spin) of the 1D chain.  
The other terms include the native Hamiltonian of the system 
\begin{equation}
  \label{eq:native-ham}
  H_{\rm S}={1\over 2}\sum_n \Delta_n^\mu \sigma_n^\mu+{1\over
    4}\sum_{n<m}J_{nm}^{\mu\nu} \sigma_n^\nu \sigma_m^\mu+\ldots
\end{equation}
describing the constant qubit couplings and the interactions between the qubits, 
and the coupling with the oscillator thermal bath,
\begin{equation}
  \label{eq:ham-env}
  H_{V}(t)=   {\textstyle\sum}_{n\mu}
  A_n^\mu\,V_n^\mu(t),\quad
  H_\sigma= {1\over2}{\textstyle\sum}_n  B_{n}^{\mu} \,  \sigma^\mu_n.
\end{equation}
In Eq.~(\ref{eq:ham-env}), $A_n^\mu\equiv A_n^\mu(p_i,q_i)$
account for the possibility of a direct coupling of the controlling
fields $V_n^\mu$ with the bath variables $q_i$, $p_i$, while
$B_n^\mu\equiv B_n^\mu(p_i,q_i)$ describe the usual coupling of the
spins with the oscillator bath.  Already in the linear response
approximation, the bath couplings~(\ref{eq:ham-env}) produce a
frequency-dependent renormalization of the control Hamiltonian $H_{\rm
  C}(t)$ [Eq.~(\ref{eq:ham-control})], as well as the thermal bath
heating via the dissipative part of the corresponding response
function.  Both effects become more of a problem with increased
spectral width of the controlling signals $V_n^\mu$.  In this work we
do not specify the explicit form of the coupling $H_V(t)$.  Instead,
we minimize the spectral width of the constructed pulses.

While the Hamiltonian~(\ref{eq:native-ham}) is a generic spin Hamiltonian, we
will also consider specifically the Hamiltonian of XXZ model with additional
on-site fields,
\begin{eqnarray}
  \label{eq:ham-xxz}
  H_{\rm S}^{XXZ}&=&   {1\over4}\sum_{\langle n,n'\rangle} \bigl[
  J_{n,n'}^z \sigma^z_n
  \sigma^z_{n'}+ 
  J_{n,n'}^\perp (\sigma^x_n \sigma^x_{n'}+\sigma^y_n
  \sigma^y _{n'})\bigr]\nonumber \\
  & & +{1\over 2}\sum_{n,\mu} 
  \Delta_n^\mu \sigma_n^\mu.
\end{eqnarray}

\subsection{Magnus expansion} 
In a 
qubit-only system with the Ha\-mil\-tonian 
\begin{equation}
  H(t)=H_{\rm C}(t)+ H_{\rm S},\label{eq:closed-ham}  
\end{equation}
the effect of the applied fields is
fully described by the evolution operator $U(t)$, 
\begin{equation}
  \label{eq:full-evolution-operator}
  U(t)\equiv T \exp\Biglb(-i\int_0^t dt'\, H(t')\Bigrb),\quad 
\end{equation}
where $T$ is the Dyson time ordering operator.

For pulses with finite width, any desired unitary transformation can
only be implemented approximately. A widely used framework to design
pulses to effect a desired unitary transformation (or, equivalently,
remove the effect of undesired terms in the Hamiltonian) is the
Magnus\cite{mehring-book} expansion and the average Hamiltonian
theory\cite{waugh-huber-haeberlen-1968,waugh-wang-huber-vold-1968}.
The expansion is done with respect to the evolution due to the control
fields alone,
\begin{equation}
  \label{eq:zeroth-evolution-operator}
  U_0(t)=T\exp\Biglb(-i\int_0^t dt'\, H_C(t')\Bigrb), 
\end{equation}
by defining the system Hamiltonian in the interaction representation (the
``rotating-frame Hamiltonian''), 
\begin{eqnarray}  
{\tilde  H}_S(t)&=&U_0^\dagger(t)H_SU_0(t).
\label{eq:interaction-representation}
\end{eqnarray}
 For a periodic control field,
$H_1(t+\tau_c)=H_1(t)$, such that the zeroth-order driven evolution is also
periodic, $U_0(t+\tau_c)=U_0(t)$, one has the following expansion 
in powers of $\tau_c$:
\begin{eqnarray}
  \label{eq:avham-expansion}
U(n\tau_c) &=& \exp(-i{\bar H} \,n\tau _c), \\
 \bar H   &=& {\bar H}^{(0)}+{\bar H}^{(1)}+{\bar H}^{(2)}+\ldots,
\end{eqnarray}
where
\begin{eqnarray}
  \label{eq:avham-zero}
  {\bar H}^{(0)}\tau_c &=&\int_0^{\tau_c}dt\, {\tilde H}_1,  \\
  \label{eq:avham-one}
  {\bar H}^{(1)}\tau_c &=&-{i\over
      2}\int_0^{\tau_c}dt_2\int_0^{t_2}dt_1\, [{\tilde 
      H}_2,{\tilde H}_1],\\
  \label{eq:avham-two}
  {\bar H}^{(2)}\tau_c & = & -{1\over 6}\int_0^{\tau_c}dt_3 
  \int_0^{t_3}dt_2\int_0^{t_2}dt_1  \\ 
  & &\times \left([{\tilde H}_3,[{\tilde
          H}_2,{\tilde H}_1]]+
    [{\tilde H}_1,[{\tilde H}_2,{\tilde H}_3]]\right).
\end{eqnarray}
Generally, the term ${\bar H}^{(k-1)}$ contains a $k$-fold integration of the
commutators of the rotating-frame Hamiltonian $\tilde H_i\equiv \tilde H_{\rm
  S}(t_i)$ at different time moments $t_i$ and scales as $\|{\bar
  H}^{(k)}\|\tau_c\propto \|\tau_c H_{\rm S}\|^{k}$.  Note that for small
enough $\tau_c$, the expansion parameter remains small even for long evolution
time.  

The Magnus expansion thus offers a basis for constructing successful
approximations towards the desired unitary evolution.  With the simpler problem
of decoupling, the goal is to have no evolution.  A $K$-th order refocusing
sequence can be defined as that where there is no evolution to $K$-th
order, that is, 
\begin{equation}
  \bar H^{(0)}=\bar H^{(1)}=\ldots =\bar H^{(K-1)}=0.
  \label{eq:refocusing-condition-K}
\end{equation}
Respectively, at
time $n\tau_c$, the error in the unitary evolution operator would scale as $\|
U(n\tau_c)-\openone\| \propto n \|\tau_c H_{\rm S}\|^{K+1}$, and the
corresponding fidelity $F(t)$ differs from unity by
\begin{equation}
  1-F(n\tau_c)\propto n^2 \|\tau_c H_{\rm S}\|^{2 K+2}.
  \label{eq:fidelity-scaling}
\end{equation}

A crucial advantage of the cumulant expansion is that the cumulants do
not contain the disconnected terms arising from different parts of the
system (cluster theorem\cite{Domb-Green-bookX,Baker-bookX}).  For an
arbitrary lattice model of the form (\ref{eq:ham-xxz}), with bonds
representing the qubit interactions, the terms contributing to $k$-th
order can be represented graphically as connected clusters involving
up to $k$ lattice bonds; for a chain of qubit such clusters cannot
have more than $n=k+1$ vertices.  Thus, to obtain the exact form of
the expansion up to and including $K$-th order, one needs to analyze
all distinct chain clusters with up to $K+1$ vertices.

\subsection{Time dependent perturbation theory}
\label{subsec:tdpt}
While the Magnus expansion is conceptually straightforward, it is
cumbersome to implement and, most importantly, the repeated
integrations are very expensive computationally already at the second
order, see Eq.~(\ref{eq:avham-one}).
An alternative strategy for evaluating high-order terms of the Magnus
expansion was suggested by the present authors in
Ref.~\onlinecite{sengupta-pryadko-ref-2005}.
Instead of working with the cumulants, the technique is based on the
time-dependent perturbation theory (TDPT).  The expansion is done
around the non-perturbed evolution due to control fields alone, see
Eq.~(\ref{eq:zeroth-evolution-operator}).  However, for actual
computations, it is more convenient to use the differential equation
\begin{equation}
  \label{eq:zeroth-equation}
  \dot U_0(t)=-iH_C(t)\,U_0(t),\quad U_0(0)=\openone.
\end{equation}
The slow evolution operator 
\begin{equation}
  \label{eq:slow-unitary}
  R(t)=U_0^\dagger(t)\, U(t).
\end{equation}
obeys the equation 
\begin{equation}
  \dot R(t)=-i \tilde H_S(t) R(t), \quad 
  \tilde H_{\rm S}(t)\equiv U_0^\dagger (t)\,   H_{\rm S}\, U_0(t),
  \label{eq:evol-oper-inter}
\end{equation}
which can be iterated to construct the standard  expansion $R(t)
  =\mathbb{I}+R_1(t) +R_2(t) +\ldots$ in powers of $(t\,H_{\rm
  S})$, 
\begin{equation}
  \label{eq:pert-expansion}
    \dot R_{k}(t)=-i \tilde H_S(t) R_{k-1}(t), \quad
    R_{0}(t)={\mathbb I}.
\end{equation}
The successive terms can be evaluated by solving, at each step, a set of
coupled first order ODE's simultaneously.  For a finite system of $n$ qubits
and a given maximum order $K$ of the expansion, one needs to solve
Eqs.~(\ref{eq:zeroth-equation}) and (\ref{eq:pert-expansion}) with $1\le k\le
K$.  These are total of $(K+1)$ coupled systems of first order ordinary
differential equations for the $2^n\times 2^n$ matrices $U_0$, $R_1$, $R_2$,
\ldots, $R_K$, and can be integrated efficiently.  Computationally, this is a 
much less challenging job
than that of evaluating repeated integrals~(\ref{eq:avham-one}),
(\ref{eq:avham-two}), and higher order terms.  For a given system,
solving the full set of equations~(\ref{eq:evol-oper-inter}) is simpler by a factor
of at least $(K+1)$.  However, it is the analysis of the perturbative
expansion that is the key for achieving the scalability of the results.

Given the set of computed $R_k(\tau_c)\equiv R_k$, the standard Magnus
expansion can be readily obtained by taking the logarithm of the
series,
\begin{eqnarray}
  \label{eq:eff-ham-gen0}
  -i H^{(0)} \tau_c &=& R_1,\\
  \label{eq:eff-ham-gen1}
  -i H^{(1)} \tau_c &=& R_2-{1\over 2}R_1^2 ,\\
  \label{eq:eff-ham-gen2}
  -i H^{(2)} \tau_c &=& R_3 -{1\over
  2}\bigl[R_1R_2+R_2 R_1\bigr] +{1\over 3}R_1^3, \,\ldots.
\end{eqnarray}
Obviously, the order-$K$
universal self-refocusing condition~(\ref{eq:refocusing-condition-K})
is formally
equivalent to
\begin{equation}
R_1(\tau_c)=R_2(\tau_c)=\ldots=R_K(\tau_c)=0.
\label{eq:self-refocusing}
\end{equation}
The matrices $R_k$ in the latter condition are much easier to evaluate
numerically using
Eqs.~(\ref{eq:zeroth-equation}), (\ref{eq:pert-expansion}).  Importantly, the
benefits of the cluster theorem are retained: to $K$-th order only clusters
with up to $K+1$ vertices need to be analyzed.

\section{Pulse design and analysis}
\subsection{Pulse design using TDPT}
The shapes of NMR-style one-dimensional 
pulses\cite{warren-herm}, self-refocusing to a given order, can be found by
analyzing 
the single-spin dynamics with the system
Hamiltonian~(\ref{eq:chemical-shift}) and the control Hamiltonian
\begin{equation}\label{eq:control-1d}
H_C(t)={1\over 2}\sigma^x V(t).
\end{equation}

Specifically, we encoded the trial pulse shapes in terms of their
Fourier coefficients,
\begin{equation}
  {V(t)} = {\phi_0\over \tau}+\Omega\sum_m A_m \cos \biglb(m\Omega
  (t-\tau/2)\bigrb), 
  \label{eq:pulse-fourier}
\end{equation}
where $\tau$ is the pulse duration, $\Omega\equiv 2\pi/\tau$ is the
corresponding angular frequency, and $\phi_0$ is the requested rotation angle
of the pulse.  Note that the form~(\ref{eq:pulse-fourier}) guarantees the
symmetry of the pulse, $V(\tau-t)=V(t)$.  In addition, in order to reduce
the spectral width of the control fields, we also constrained a certain number
of derivatives of the function~(\ref{eq:pulse-fourier}) to vanish at $t=0$ and
$t=\tau$, $V^{(l)}(0)=0$, $l=0,1,\ldots, 2L-1$.

We implemented the computational algorithm  described in the
previous section using the
standard fourth-order Runge-Kutta algorithm for solving coupled
differential equations ~(\ref{eq:zeroth-equation}),
(\ref{eq:pert-expansion}), and the GSL library\cite{gsl-ref} for
matrix operations.  The coefficient optimization was done using a
combination of simulated annealing and the steepest descent method.
The target function for single-pulse optimization included the sum of
the magnitudes squared of the matrix elements of the matrices
$R_k\equiv R_k(\tau)$, $k=1,\ldots,K$, as well as the weighed sum of the squares
of the coefficients $A_m$,
\begin{equation}
  \label{eq:target-fun}
  f_K=\left(\sum_{k=1}^K \tr R_k^\dagger R_k\right)^{1/2} +\epsilon
  \sum_{m=1}^{M} 
  m^2 A_m^2  .
\end{equation}
The second sum serves to provide some
suppression of the higher Fourier harmonics of the pulse.
In our simulations, the minimization was considered as having converged only
after the first term evaluated to zero with numerical precision
(typically, eight digits or better).  For such a minimum to exist, the
coefficient $\epsilon$ in Eq.~(\ref{eq:target-fun}) should be sufficiently
small (we used $\epsilon=10^{-4}$).  

For given pulse order $K$ and the given number of additional
constraints $L$, there is a minimum number of harmonics
$M_\mathrm{min}(K,L)$ necessary for convergence.  However, we found
that the shapes obtained with $M=M_\mathrm{min}(K,L)$ tend be
over constrained and simply do not look nice.  Our solution was to add
one or two additional Fourier harmonics by increasing $M$.

\subsection{Pulse shapes}

Previously, in Ref.~\onlinecite{sengupta-pryadko-ref-2005}, we gave
the coefficients of the first-order self-refocusing ($K=1$) inversion
($\phi_0=\pi$) pulse shapes $S_L$, as well as the second-order ($K=2$)
inversion shapes $Q_L$, $L=1,2$.  Here $L$ is the parameter that
indicates the number of constraints at the ends of the interval: the
value of the function and its derivatives up to $(2L-1)$st vanish at
the ends of the interval [note that all odd derivatives are suppressed
  automatically due to the symmetry of the function, see
  Eq.~(\ref{eq:pulse-fourier})].

In this work, we extend the list of constructed pulses to rotation
angles $\phi_0=10^\circ$, $20^\circ$, \ldots, $180^\circ$.  In Table
~\ref{tab:coeff}, we list the coefficients for the pulses used in the
simulations.  The coefficients for all of the constructed pulses are
available upon request. 

\begin{table*}[htb]
  \centering
  \begin{tabular}[c]{c|c|c|c|c|c|c}
    & $A_0$& $A_1$ & $A_2$ & $A_3$ & $A_4$ & $A_5$ \\
    \hline 
    $S_1(2\pi)$ &  1.0&  -0.0237996956&  -0.6226198703 & -0.3535804341\\
    $S_2(2\pi)$ &  1.0&  -0.0294359406&  -1.1741824154 & -0.2097531295&   0.4133714855\\
    $Q_1(2\pi)$ &  1.0&   2.1406171699&  -2.3966480505&  -0.6474844418 & -0.0964846776\\
    $Q_2(2\pi)$ &  1.0&   1.4818894659&  -2.6971749102&
    -0.4384679067&   0.3434236044 &  0.3103297466\\ 
    \hline 
    $S_1\equiv S_1(\pi)$ & 0.5 &  -1.2053193822 &  0.4796467863 &    0.2256725959 & \ \\
    $S_2\equiv S_2(\pi)$ & 0.5 &  -1.1950692860 &   0.7841592117 &     0.0737326786 &  -0.1628226043 \\
    $Q_1\equiv Q_1(\pi)$ & 0.5 &   -1.1374072085 &   1.5774920785&   -0.6825355002&  -0.2575493698\\
    $Q_2\equiv Q_2(\pi)$ &0.5 &   -1.0964843348 &  1.5308987822&
    -1.1472441408&   0.0025173181 &  0.2103123753\\
    \hline  
    $ S_1(\pi/2)$ &  0.25 &  -1.8963102551 &  1.1337663752 &  0.5125438801 & \ \\
    $ S_2(\pi/2)$ & 0.25 &  -1.9049987341 &   1.9858884053 &   0.1063314501 & -0.4372211211\\
    $ Q_1(\pi/2)$ &  0.25 &  -1.8948543589 &   0.5873324062 &   0.5970352560 &   0.4604866969\\
    $ Q_2(\pi/2)$ & 0.25 & -2.1145695246 &  0.6415685732 &   1.6854185871 &   0.4511145740 & -0.9135322049\\
    \hline               
  \end{tabular}
  \caption{Fourier coefficients for the constructed pulses, see
    Eq.~(\ref{eq:pulse-fourier}).  Shapes $S_L(\phi_0)$ and $Q_L(\phi_0)$ are
    the pulse shapes for rotation angle $\phi_0$, 
    respectively first  ($K=1$) and second ($K=2$) order for the
    Hamiltonian~(\protect\ref{eq:chemical-shift}).  These shapes 
    have $2L$  derivatives vanishing at the ends of the interval.} 
  \label{tab:coeff}
\end{table*}

\begin{figure}[htp]%
  \centering
  \epsfxsize=0.7\columnwidth
  (a) \epsfbox{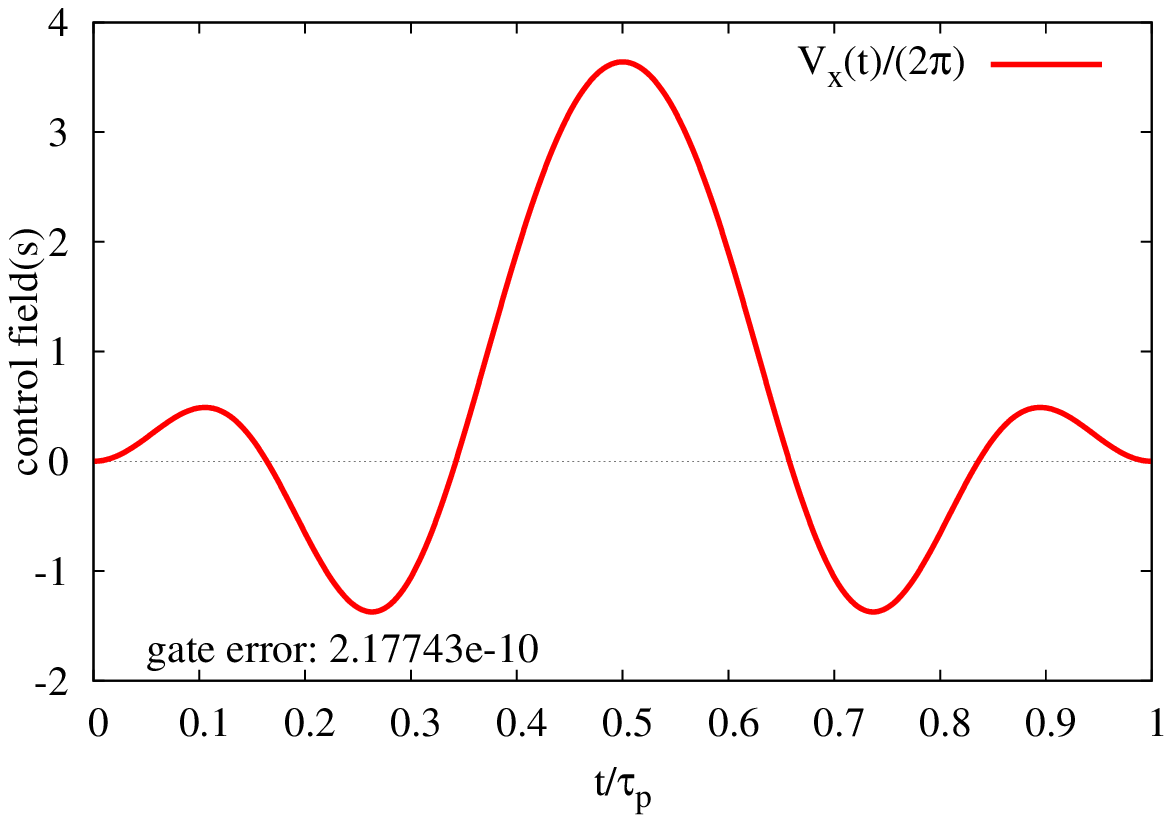}\\
  \epsfxsize=0.7\columnwidth
  (b) \epsfbox{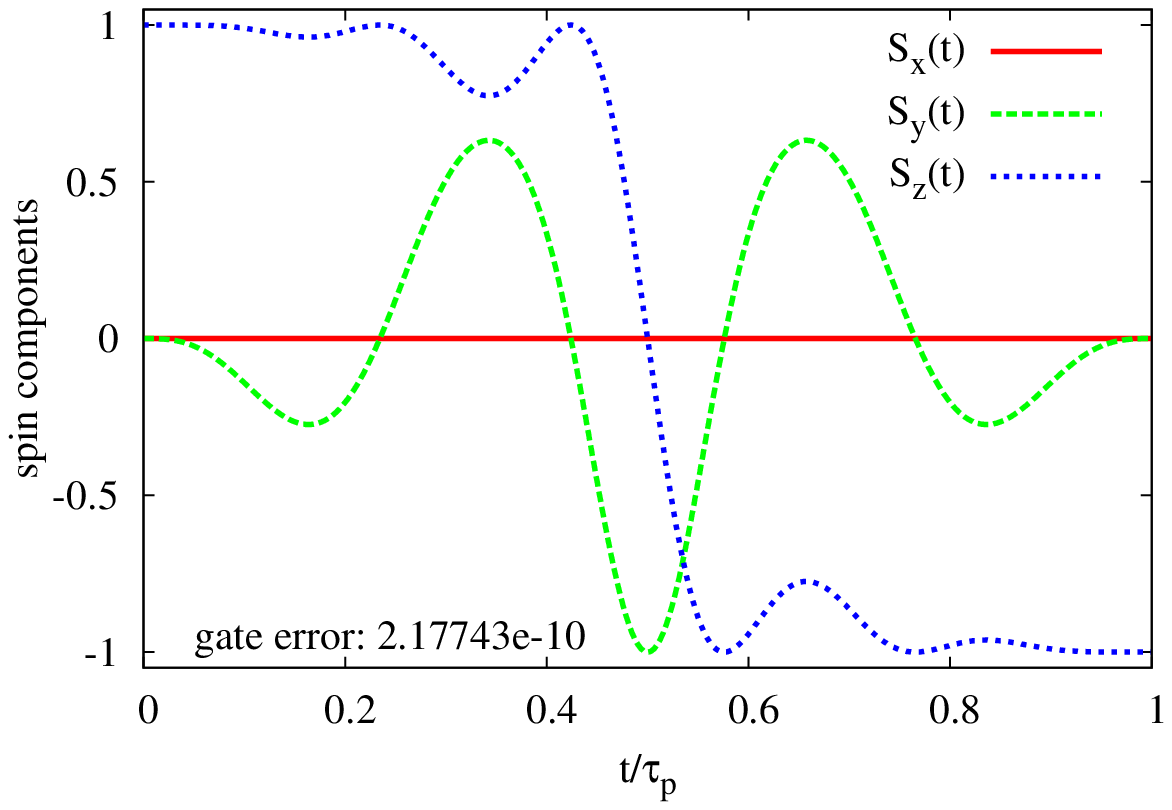}\\
  \epsfxsize=0.7\columnwidth
  (c) \epsfbox{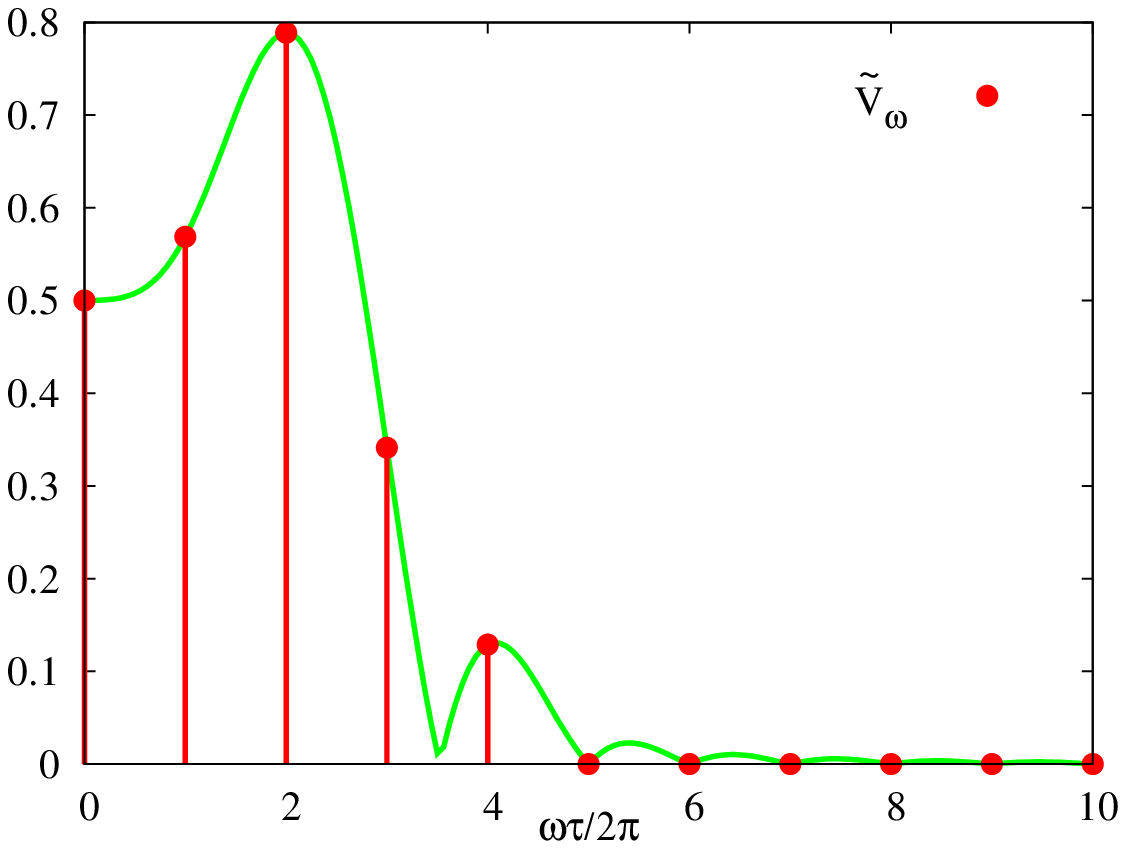}
     \caption{A single-spin second-order inversion ($\pi$) pulse $Q_1(\pi)$. 
       (a) Pulse profile over a complete period, (b) Evolution of the
       spin beginning with $s_z=1$. (c)
       The power spectrum of the 
       pulse. The vertical lines denote the location of the harmonics.
       As seen, the spectral weight is almost entirely confined to
       $\omega < 5\omega_0$.}
     \label{fig:pulseQ1x180}
\end{figure}

\begin{figure}[htp]%
  \centering
  (a) \includegraphics[width=0.7\columnwidth]{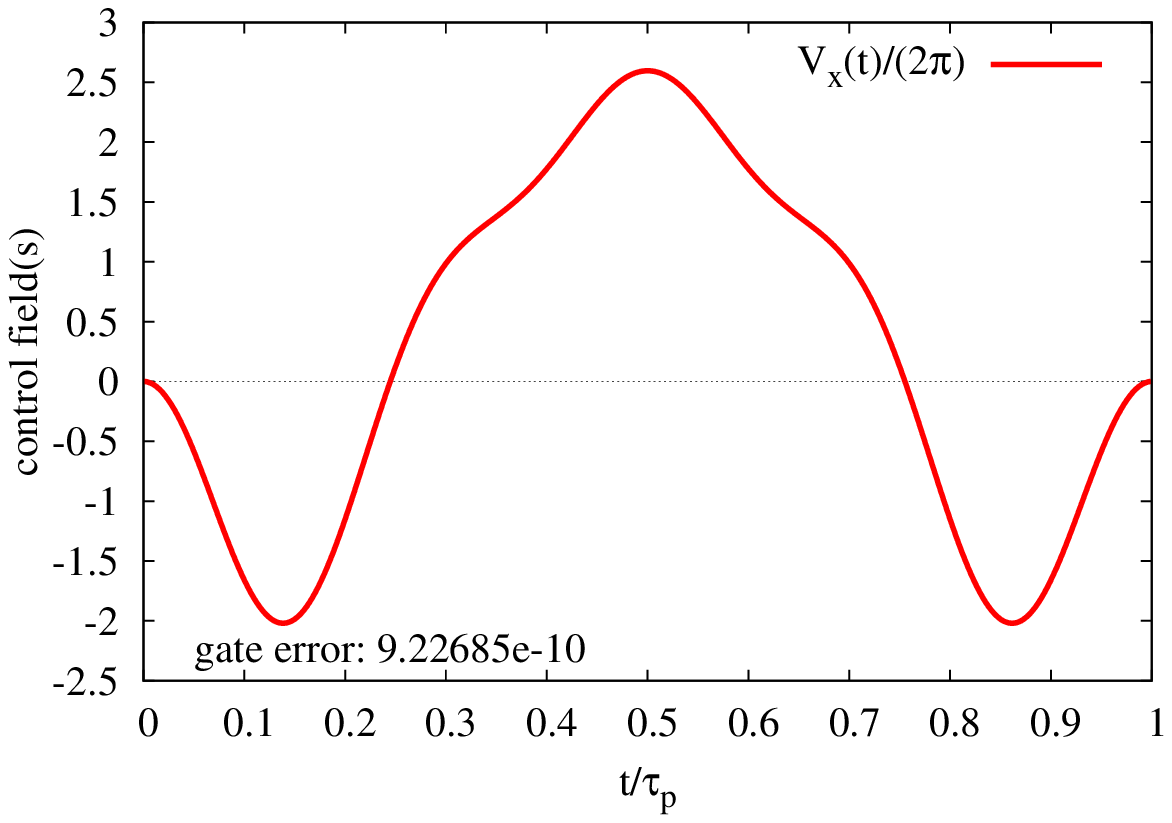}\\
  (b) \includegraphics[width=0.7\columnwidth]{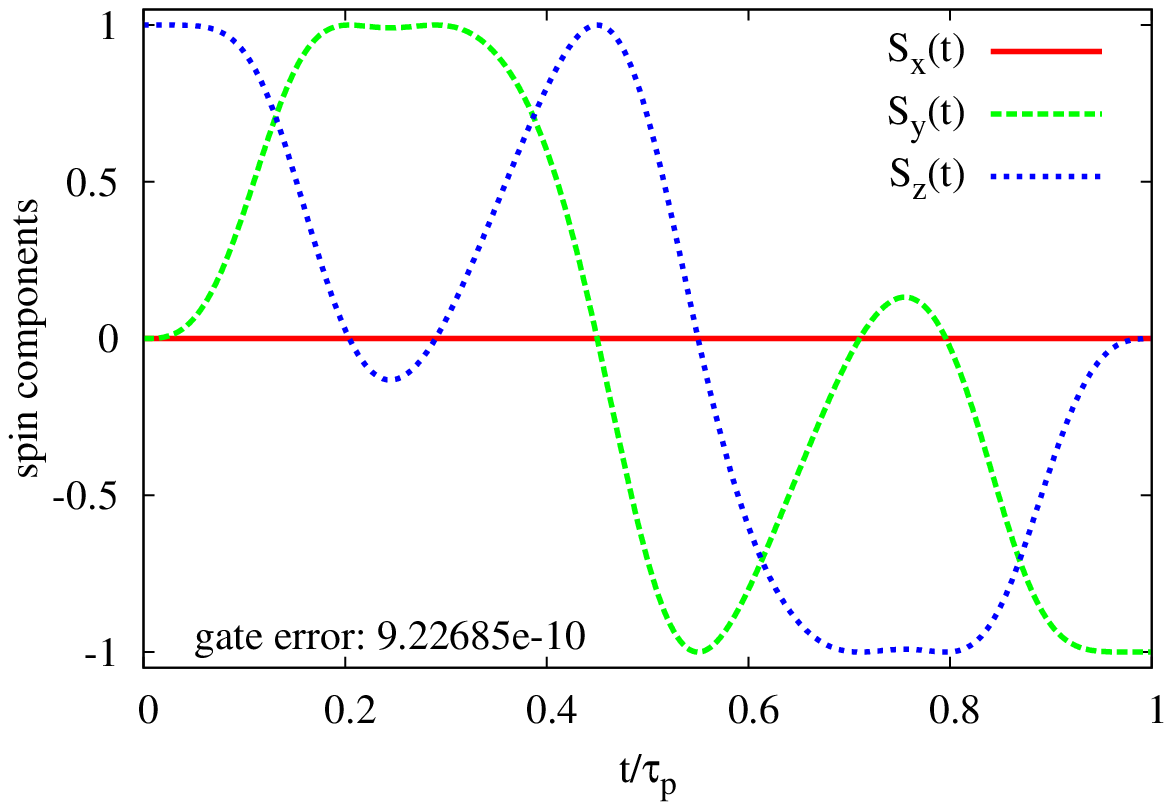}\\
  (c) \includegraphics[width=0.7\columnwidth]{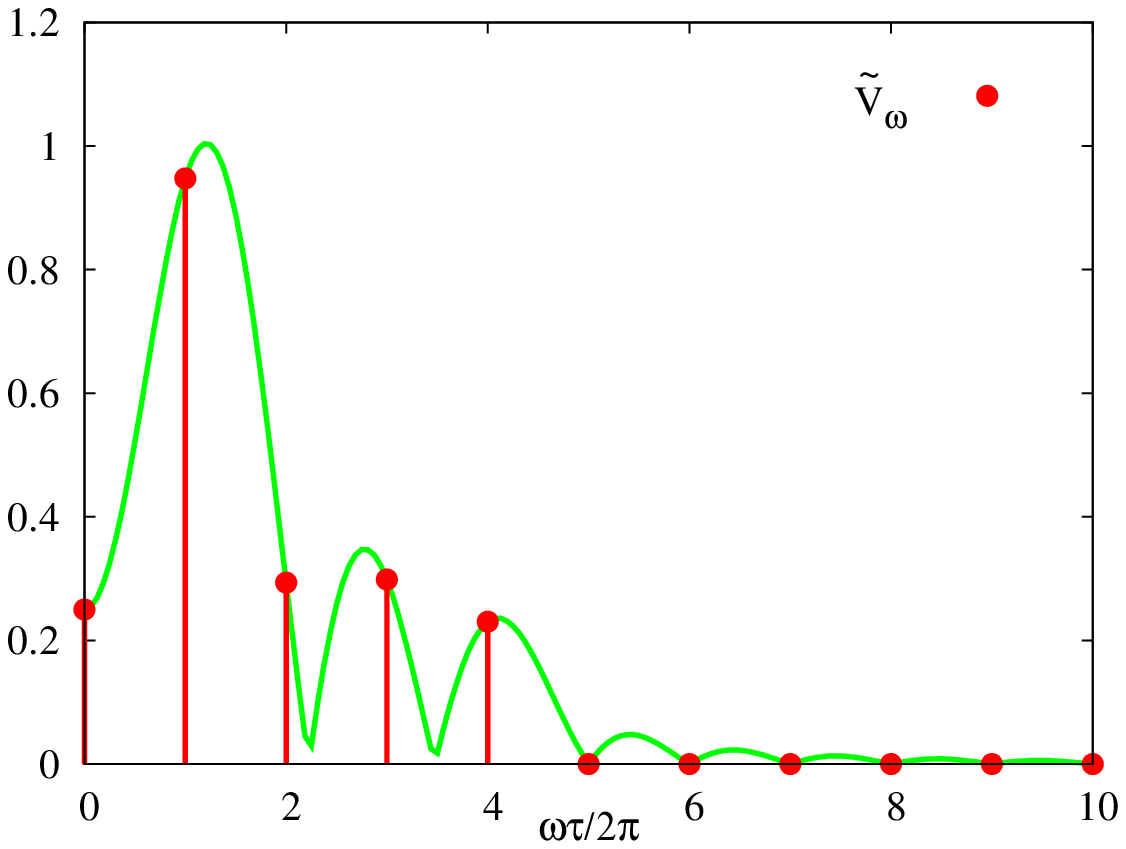}
     \caption{As in Fig.~\ref{fig:pulseQ1x180}, but for the
  single-spin second-order $\pi/2$ pulse $Q_1(\pi/2)$.  }
     \label{fig:pulseQ1x090}
\end{figure}
\begin{figure}[htbp]
  \centering
    \epsfxsize=0.9\columnwidth
 \epsfbox{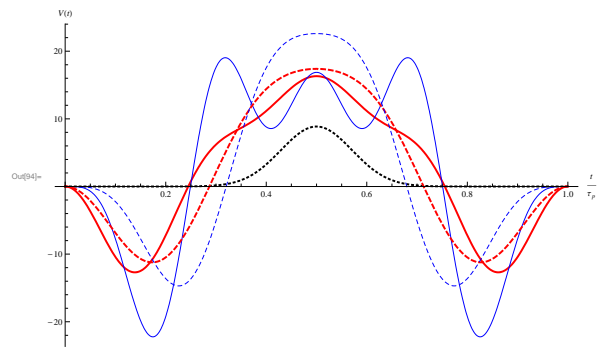}
  \caption{(color online) Pulse shapes for $\phi_0=\pi/2$.  Solid lines
    represent 
    $Q_L(\pi/2)$, dashed lines correspond to $S_L(\pi/2)$.  Pulse
    shapes with $L=1$ are drawn with thin blue lines, while those with
    $L=2$ are drawn with thick red lines.  The black doted line shows
    the Gaussian shape $G_{010}(\pi/2)$.}
  \label{fig:90deg}
\end{figure}
\begin{figure}[htbp]
  \centering
    \epsfxsize=0.9\columnwidth
 \epsfbox{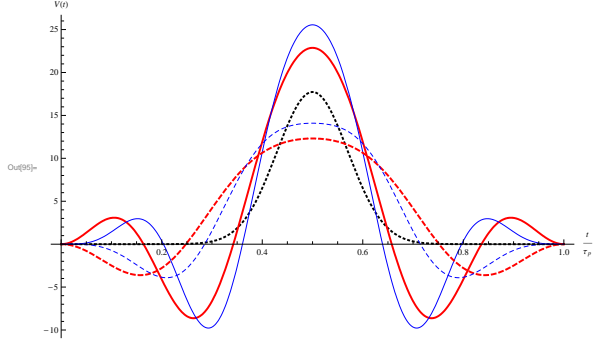}
  \caption{(color online) As in Fig.~\ref{fig:90deg} but for the
    inversion pulses, $\phi_0=\pi$.  Note that the 1st order pulses
    ($S_L(\pi)$, dashed lines) actually have a smaller power than the
    Gaussian pulse.}
  \label{fig:180deg}
\end{figure}
\begin{figure}[htbp]
  \centering
    \epsfxsize=0.9\columnwidth
 \epsfbox{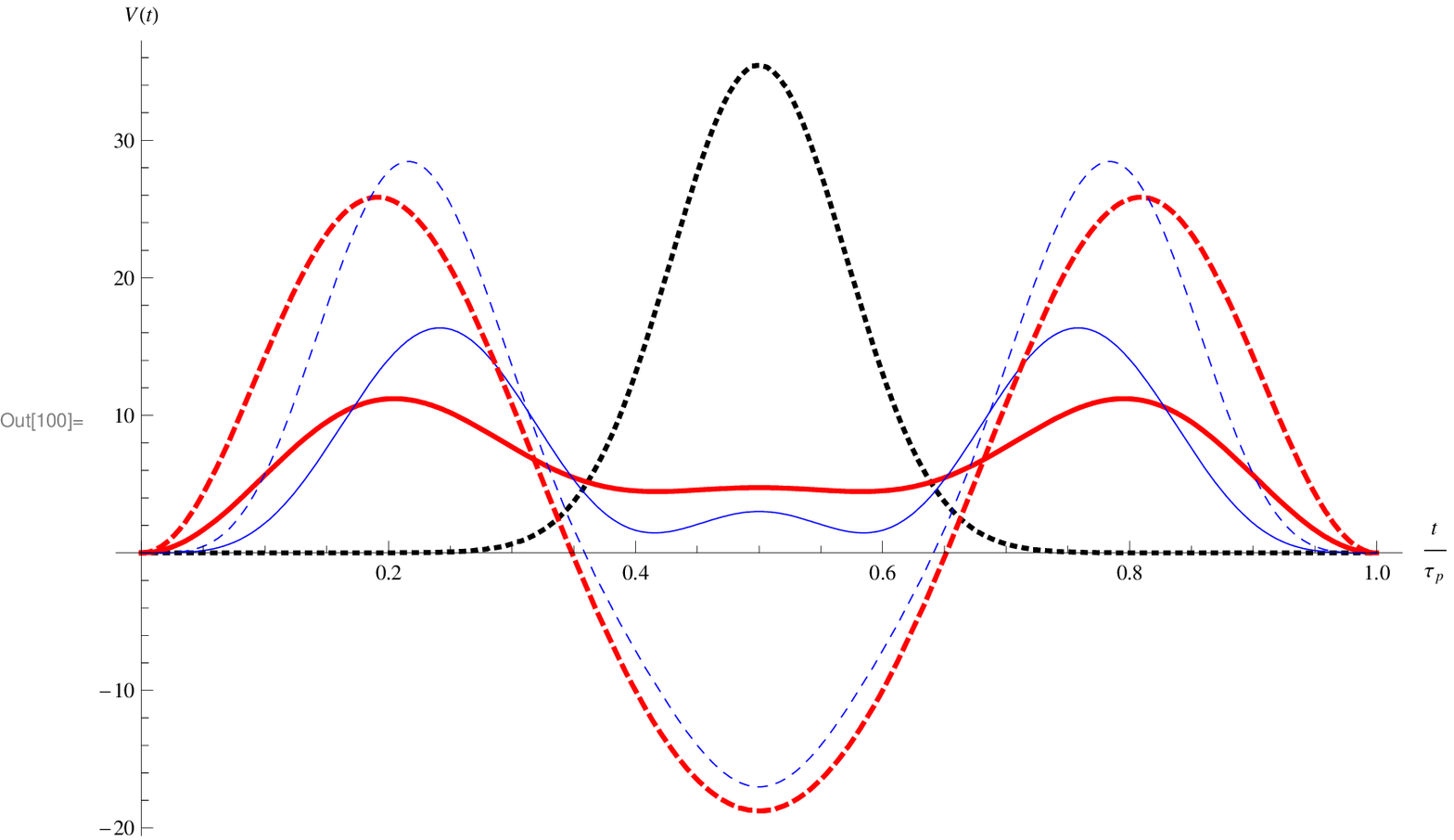}
  \caption{(color online) As in Fig.~\ref{fig:90deg} but for the
    pulses with $\phi_0=2\pi$.  The pulse shapes appear to resemble
    those of a pair of consecutive $\pi$ pulses.  Second-order pulses
    happen to have the smallest power.}
  \label{fig:360deg}
\end{figure}
\subsection{Pulse shape analysis}
The pulse shapes $Q_L(\phi_0)$, $S_L(\phi_0)$ are constructed as first-
or second-order self-refocusing pulses for the chemical shift
Hamiltonian, Eq.~(\ref{eq:chemical-shift}).  We would like, however,
to have a universally good pulse shape that would work in most
settings.  To analyze the performance of the constructed pulses in
most general circumstances, we construct the Magnus expansion of the
evolution operator for the most general system Hamiltonian,
\begin{equation}
  \label{eq:ham-system-general}
  H_S=A_0+A_x\sigma^x+A_y\sigma^y+A_z\sigma^z,
\end{equation}
where $A_\mu$ are the operators responsible for coupling with the outside
worlds, and $A_0$ is the external Hamiltonian.  The analysis of the inversion
pulses with $\phi_0=\pi$ appeared previously in
Ref.~\onlinecite{pryadko-quiroz-2007}; here we extend it to $\phi_0\neq \pi$.

\subsubsection{Driven evolution}
The control Hamiltonian (\ref{eq:control-1d}) alone [to zeroth
order in $H_S$] produces the  following unitary evolution
operator [cf.~Eq.~(\ref{eq:zeroth-evolution-operator})] 
\begin{equation}
  U_0(t)=e^{-i\sigma^x \phi(t)/2}, \quad \phi(t)\equiv \int_0^t
  dt'\,V(t').\label{eq:control-evolution}   
\end{equation}
When acting on the spin operators, this is just a rotation,
$U_0^\dagger(t)\sigma^yU_0 (t)=\sigma^y\cos \phi(t)-\sigma^z \sin
\phi(t)$.  Consequently, the system Hamiltonian in the interaction
representation has the form
\begin{eqnarray}
  \tilde H_S(t)&=& A_0+\sigma^x A_x+
  \sigma^y (A_y \cos\phi+A_z\sin\phi)\nonumber\\ 
  & & +\sigma^z (A_z \cos\phi-A_y\sin\phi).
  \label{eq:ham-interaction-representation}
\end{eqnarray}
\subsubsection{Leading-order average Hamiltonian}
The zeroth order average Hamiltonian~(\ref{eq:avham-one}) is just the
average of Eq.~(\ref{eq:ham-interaction-representation}) over the pulse
duration.  We assume $V(t)$ represents a symmetric pulse,
$V(\tau_p-t)=V_x(t)$.  Then, $\phi(t)$ is antisymmetric,
$\phi(\tau_p-t)=\phi_0-\phi(t)$, where $\phi_0\equiv\phi(\tau_p)$ is the
overall notation angle.  It is convenient to introduce the symmetrized
rotation angle, 
\begin{equation}
\varphi(t)\equiv \phi(t)-\phi_0/2,\label{eq:symmetrized-angle}
\end{equation}
such that
$\varphi(\tau_p-t)=-\varphi(\tau_p)$.  Then, the average  of the sine
over the pulse duration vanishes, $\langle \sin\varphi\rangle  =0$.
This implies that the 
averages of the cosine and sine of  the original rotation angle are
\begin{eqnarray}  
\label{eq:average-C}
\C&\equiv &\langle\cos\phi\rangle =\cos(\phi_0/2)\langle \cos
\varphi\rangle , \\
\S&\equiv& \langle\sin\phi\rangle =\sin(\phi_0/2)\langle \cos \varphi\rangle ,
\label{eq:average-S}
\end{eqnarray}
where 
\begin{equation}
  \label{eq:pulse-average}
  \langle f(t)\rangle \equiv {1\over \tau_p}\int_0^{\tau_p} dt\,f(t) .
\end{equation}
If we denote 
\begin{equation}
  \upsilon\equiv \langle \cos\varphi\rangle 
  =\int_0^{\tau_p} {dt\over \tau_p}\cos\varphi(t),
  \label{eq:upsilon-defined}
\end{equation}
then the zeroth order
average Hamiltonian for a one-dimensional pulse becomes
\begin{eqnarray}
  \bar H^{(0)}&=& A_0+\sigma^x A_x+ \upsilon \Bigr[
  \sigma^y \Bigl(A_y \cos{\phi_0\over 2}+A_z\sin{\phi_0\over2}\Bigr)\nonumber\\ 
  & & +\sigma^z \Bigl(A_z \cos{\phi_0\over2}-A_y\sin{\phi_0\over2}\Bigr)\Bigr].
  \label{eq:zeroth-order-ham}
\end{eqnarray}
For the special case of the chemical-shift
Hamiltonian~(\ref{eq:chemical-shift}), we have $A_0=A_x=A_y=0$,
$A_z=\Delta/2$, and Eq.~(\ref{eq:zeroth-order-ham}) gives 
\begin{equation}  
  \bar H^{(0)}= \upsilon {\Delta\over 2}\left( \sigma^z
    \cos{\phi_0\over2}+\sigma^y\sin{\phi_0\over2}  \right).
  \label{eq:zeroth-order-ham-cs}
\end{equation}
Clearly, the 1st-order self-refocusing condition corresponds to
$\upsilon=0$.  For such pulses the full zeroth-order average 
Hamiltonian is given just by the two first terms in
Eq.~(\ref{eq:zeroth-order-ham-cs}).

\subsubsection{1st-order average Hamiltonian}
The 1st-order average Hamiltonian~(\ref{eq:avham-one}) is given by a double
integral of the commutator of the system Hamiltonian in the interaction
representation (\ref{eq:ham-interaction-representation}) evaluated at two
different times.  We note that every term in
Eq.~(\ref{eq:ham-interaction-representation}) can be classified as either
time-independent [proportional to $e(t)\equiv 1$], proportional to $c(t)\equiv
\cos\phi(t)$, or to $s(t)\equiv\sin\phi(t)$.  Therefore, most generally, the
second-order terms in the evolution operator can contain the following nine
integrals,
\begin{eqnarray}
  \label{eq:2nd-order-coeff}
  \EE\equiv \langle 1'1\rangle={1\over2},&\;&\!\!
  \EC\equiv \langle 1'\cos\phi\rangle,\quad
  \ES\equiv \langle 1'\sin\phi\rangle,\nonumber\\
  &\;&\!\! \CE\equiv\langle \cos\phi'\,1\rangle,\quad 
  \SE\equiv\langle \sin\phi'\,1\rangle, \nonumber\\
  \CC\equiv\langle \cos\phi'\cos\phi\rangle ,&\;&\!\!    
  \CS\equiv\langle \cos\phi'\sin\phi\rangle,\nonumber\\     
  \SC\equiv\langle\sin\phi'\cos\phi\rangle,&\;&\!\!    
  \SSS\equiv\langle\sin\phi'\sin\phi\rangle,
\end{eqnarray}
where we used the notation
\begin{equation}
  \langle f(\phi') g(\phi)\rangle\equiv {1\over \tau^2}\int_0^\tau dt'\,
  f\biglb(\phi(t')\bigrb)
  \int_0^{t'} dt\, g\biglb(\phi(t)\bigrb), \label{eq:double-average}
\end{equation}
and $1\equiv e(t)$ or $1'\equiv e(t')$ indicate an identity factor at the
corresponding position of the average.  However, because of the commutator
structure in Eq.~(\ref{eq:avham-one}), only the following antisymmetric
combinations appear in the expression for the corresponding term in the
average Hamiltonian theory, $H^{(1)}$,
\begin{equation}
  \label{eq:2nd-order-defs}
{\alpha\over2}\equiv {\SC-\CS\over2}, \quad 
\zeta_C\equiv {\EC-\CE\over2},\quad 
\zeta_S\equiv {\ES-\SE\over2}, 
\end{equation}
where 
\begin{equation}
  \label{eq:alpha-defined}
  \alpha\equiv {1\over2\tau_p^2}\int_0^{\tau_p} dt' \int_0^{t'} dt\,
  \sin\biglb(\phi(t')-\phi(t)\bigrb)
\end{equation}
In fact, the coefficients $\zeta_C$ and $\zeta_S$ can be reduced further, 
\begin{equation}
  \label{eq:zetas-defined}
  \zeta_C=\zeta\sin{\phi_0\over2},\quad 
  \zeta_S=-\zeta\cos{\phi_0\over2},
\end{equation}
where [see Eq.~(\ref{eq:symmetrized-angle}) for the definition of
   $\varphi(t)$]
\begin{equation}
  \zeta\equiv \int_0^{\tau_p}
  {dt'\over \tau_p}\Bigl({t\over \tau_p}-{1\over2}\Bigr)
  \sin\varphi(t). 
  \label{eq:zeta-defined}
\end{equation}
Thus, to second order, the average Hamiltonian of a symmetric angle-$\phi_0$
one-dimensional pulse is determined by only three dimensionless coefficients,
$\upsilon$, $\alpha$, and $\zeta$, see Eqs.~(\ref{eq:upsilon-defined}),
(\ref{eq:alpha-defined}), and (\ref{eq:zeta-defined}).  These coefficients
contain all the relevant information about the shape of the pulse.

An explicit calculation of  the 1st-order average Hamiltonian gives 
\begin{widetext}
\begin{eqnarray*}
H^{(1)}&=&  \alpha\tau_p (i  [A_z,A_y]-\sigma^x (A_y^2+A_z^2))\\ 
 & +& 
 \zeta\tau_p\cos {\phi_0\over2}
\Biglb(\sigma^y (i  [A_z,A_0]+ \{A_x,A_y\})+\sigma^z(i  [A_0,A_y]+ \{A_x.A_z\})\Bigrb)\\ 
& -&  \zeta\tau_p\sin {\phi_0\over2}
\Biglb(\sigma^y(i  [A_y,A_0]- \{A_x,A_z\})+\sigma^z(i  [A_z,A_0]+ \{A_x,A_y\})\Bigrb).\hskip3in
\end{eqnarray*}
\end{widetext}
For the Hamiltonian (\ref{eq:chemical-shift}), the terms with $\zeta$
disappear, and we have, simply
\begin{equation}
  \label{eq:ham-chem-1}
  H^{(1)}= - \alpha   \sigma_x {\Delta^2\tau_p\over 4}.
\end{equation}
Thus, the second-order self-refocusing pulses have both $\upsilon=0$ and
$\alpha=0$.  

The actual parameters for the pulses with $\phi_0=\pi/2$, $\pi$, and $2\pi$
are listed in TAB.~\ref{tab:params}.

\begin{table}[htc]
  \centering
  \begin{tabular}[c]{c|c|c|c|c}
    pulse & $\phi_0$ & $\upsilon\equiv \langle \cos\varphi\rangle $ &
    $\alpha$ & $\zeta$      \\[0.05in]\hline
    $\phi_0\delta(t-\tau_p/2)$ & $\phi_0$ &
    $\displaystyle\cos{\phi_0\over2}$ &
    $\displaystyle{\sin\phi_0\over4}$ & $\displaystyle
    {1\over4}\sin{\phi_0\over2}$ \\ [0.1in]
    \hline
    $\displaystyle{\pi\over2}^{\strut}\delta(t-\tau_p/2)$ & $\pi/2$ & $\sqrt2/2$ & $1/4$ & $\sqrt2/8$\\
    $G_{0.05}[90]$ & ${\pi }/{2}$ & $0.730111$ & $0.398519$ & $0.175999$\\
    $G_{0.1}(90)$ & ${\pi /2}$ & $0.753116$ & $0.420275$ & $0.173665$\\
    $S_1(90)$& $\pi/2$ & $0$ & $-0.013067$ & $0.198719$\\
    $S_2(90)$& $\pi/2$ & $0$ & $-0.0294665$ & $0.182109$\\
    $Q_1(90)$& $\pi/2$ & $0$ & $0$ & $0.202067$\\
    $Q_2(90)$& $\pi/2$ & $0$ & $0$ & $0.161658$\\[0.05in]
    \hline 
    $\displaystyle{\pi}(t-\tau_p/2)$ & $\pi$ & $0$ & $0$ & $1/4$\\
    $G_{0.05}(180)$ & $\pi $ & $0.0744894$ & $0.0377451$ & $0.249476$\\
    $G_{0.1}(180)$ & $\pi $ & $0.148979$ & $0.0764911$ & $0.247905$\\
    $S_1(180)$& $\pi $ & $0$ & $0.0332661$ & $0.238227$\\
    $S_2(180)$& $\pi $ & $0$ & $0.0250318$ & $0.241378$\\
    $Q_1(180)$& $\pi $ & $0$ & $0$ & $0.239888$\\
    $Q_2(180)$& $\pi $ & $0$ & $0$ & $0.242209$\\[0.05in]    
    \hline
    $\displaystyle{2\pi}(t-\tau_p/2)$ & $2\pi$ & $-1$ & $0$ & $0$\\    
    $G_{0.05}(360)$ & $2 \pi $ & $-0.896959$ & $0.402852$ & $0.00291436$\\
    $G_{0.1}(360)$ & $2 \pi $ & $-0.793918$ & $0.317488$ & $0.0116577$\\
    $S_1(360)$& $2 \pi $ & $0$ & $0.0739621$ & $0.113233$\\
    $S_2(360)$& $2 \pi $ & $0$ & $0.0612747$ & $0.0811486$\\    
    $Q_1(360)$& $2 \pi $ & $0$ & $0$ & $0.00403872$\\
    $Q_2(360)$& $2 \pi $ & $0$ & $0$ & $0.00734526$\\\hline
  \end{tabular}
  \caption{Parameters of several common pulse shapes.  The first line
    represents  the ``hard'' $\delta$-function pulse applied at the center of
    the interval of duration $\tau$, $G_{001}$ denotes the 
    Gaussian pulse with the width $0.01\tau_p$, while $S_n$ and $Q_n$ denote
    the 1st and 2nd-order self-refocusing pulses from  Tab.~\ref{tab:coeff}.} 
  \label{tab:params}
\end{table}

\section{Open systems}
\label{sec:open}
In this work we concentrate on the performance of high-order pulses and pulse
sequences in closed quantum systems.  However, it turns out that such
sequences also remain efficient in open systems, in the presence of
low-frequency bath
modes\cite{pryadko-sengupta-kinetics-2006,pryadko-quiroz-2007}.  

The analysis is done in general form with the help of an assumption that the
bath couplings have the same form as the existing terms in the system
Hamiltonian~(\ref{eq:native-ham}), which are assumed to be suppressed to order
$K=1$ or $K=2$.  The bath modes are assumed to be low-frequency; in addition
to the expansion in powers of the corresponding couplings, one needs a low
frequency expansion in powers of the \emph{adiabaticity parameter}
$\tau_c/\tau_0$, where $\tau_c$ is the decoupling cycle duration and $\tau_0$
is the bath correlation time.  

With $K=1$ decoupling, the effect in the open system is a suppression of
direct decay ($T_1$) processes, as well as the reduction of the dephasing rate
($T_2$) by the factor of order of the adiabaticity parameter $\tau_c/\tau_0$.
The former result can be understood by analyzing the spectral properties of
the driven system\cite{kofman-kurizki-2001,kofman-kurizki-2004}, while the
latter can be viewed as due to a reduction of the time step for phase
diffusion.  With second-order decoupling, $K=2$, the decoherence rate is
additionally suppressed, and with time-reversal invariant bath coupling all
orders of the expansion in powers of adiabaticity parameter may vanish, in
which case the leading-order dephasing term becomes exponentially small and
dephasing would likely be determined by terms of higher order in bath coupling.
Along with the decoherence rates characterizing the exponential decay of
quantum correlations with time, the corresponding prefactor, which determines
the ``visibility'' (or ``initial decoherence''\cite{facchi-nakazato-2004}),
was also analyzed\cite{pryadko-sengupta-kinetics-2006}.  While for generic
refocusing sequences with $K\ge1$ the initial decoherence is quadratic in
$\tau_c$ and does not scale with the thermal bath correlation time $\tau_0$,
for symmetric pulse sequences it is reduced by an additional power of the
adiabaticity parameter $(\tau/\tau_0)$.  These results were originally derived
for a generic featureless bath, but they also hold in a vicinity of a sharp
resonance as long as the {\em effective\/} (i.e., renormalized as in the
average Hamiltonian) coupling to the corresponding mode is small compared to
its width\cite{pryadko-quiroz-2007}.

\section{Application examples.}
\subsection{Decoupling sequences for a chain of qubits}
Decoupling sequences are designed to prevent quantum evolution from happening.
Thus, we want to construct a sequence such that the resulting evolution
operator over the period $\tau_c$ is identity, $U(\tau_c)=\openone$.  We
illustrate the \emph{scalability} of dynamical decoupling to large system
sizes by considering linear chains of qubits with either Ising or XXZ n.n.\
random-valued couplings [only $J_{n,n+1}^z$ or both $J_{n,n+1}^z$ and
$J_{n,n+1}^x=J_{n,n+1}^y$ in Eq.~(\ref{eq:native-ham})], plus the local fields
either along $z$ axis or in arbitrary direction [$\Delta^z_n\neq0$ or
$\Delta^\mu_n\neq0$ for $\mu=x,y,z$ in Eq.~(\ref{eq:native-ham})].

With such a system Hamiltonian, zeroth-order average
Hamiltonian~(\ref{eq:avham-zero}) contains only individual qubits or pairs of
neighboring qubits, the largest clusters contributing to the 1st-order average
Hamiltonian~(\ref{eq:avham-one}) originate from two bonds sharing a site
(three qubits), and in general $\bar H^{(n)}$ contains terms spanning
contiguous clusters of up to $n+1$ bonds, that is, $n+2$ qubits.  Thus, to
design a $K$-th order decoupling sequence, one needs to consider individual
clusters of up to $K+1$ qubits.

With nearest-neighbor and local couplings only, the decoupling can be
implemented by simultaneously applying pulses on either odd or even
sublattice.  We note that in our setup there is no gap between subsequent
pulses, the pulses follow back to back with the repetition period $\tau$.  The
system is ``focused'' at the end of each cycle consisting of several pulses of
length $\tau$.  Such a scheme with a common ``clock'' time $\tau$ is
convenient, e.g., for parallel execution of quantum gates in different parts
of the system.  For each qubit, various pulses (or intervals of no signal) can
be executed in sequence.

In this work we consider the following two sequences from
Ref.~\onlinecite{sengupta-pryadko-ref-2005},
$\mathbf{4}=\X_1\Y_2\bar\X_1\bar\Y_2$ and its symmetrized version $\mathbf{8}=
\X_1\Y_2\bar\X_1\bar\Y_2\bar\Y_2\bar\X_1\Y_2\X_1$, which provide universal
refocusing of the couplings between the sublattices, and also suppress the
on-site chemical shifts $\Delta_n^z$.  Here, $\X_1$ is a $\pi_x$ pulse
simultaneously applied on all odd sites, $\bar\Y_2$ is a $(-\pi)_y=\pi_{-y}$
pulse applied on all even sites, etc.  These sequences are ``best'' sequences
at given length for all pulse shapes found by exhaustive search (high-order
sequences\cite{brown-harrow-chuang-2004,khodjasteh-Lidar-2005} equivalent for
hard pulses do not necessarily have equal orders here).  The fact that such a
brute-force optimization approach works is entirely due to the efficiency of
the numerical method.

In addition, we constructed two longer sequences,
$\mathbf{16}=\X_1\Y_2\Y_1 0
\bar\X_1\X_2\Y_10\X_1\bar\Y_2\Y_10\bar\X_1\X_2\Y_10$, 
and its symmetrized version $\textbf{32}$, constructed by running the sequence
$\textbf{16}$ first directly and then in reverse order.  Here $0$ denotes
zero pulse, an empty interval of  duration $\tau$.  These two sequences provide
universal decoupling both for any couplings between the sublattices and for
arbitrary on-site fields ($\Delta_n^\mu\neq0$).

In addition to the system Hamiltonian, the effectiveness of a sequence
application depends on the quality of the pulses.  In
Table~\ref{tab:decoup-order}, we list orders of the sequences when applied
with different pulse shapes, computed using the numerical time-dependent
perturbation theory as described in sec.~\ref{subsec:tdpt}.  The term
$R_k(\tau_c)$ was considered to be zero if its norm vanished with numerical
precision, typically $10^{-8}$ or better, compared to typical values of order
one for orders where $R_k(\tau_c)\neq0$.   The orders $K$ do not depend
on the chain length; we verified this statement on chains up to $n=7$ qubits.
Also, the computed orders are the same for all self-refocusing pulse shapes of
particular order; we believe that the results will remain valid for other
symmetric pulse shapes of the same order as indicated in the 1st column of
Table~\ref{tab:decoup-order}.

\begin{table}[htbp]
  \centering
  \begin{tabular}[c]{c|c|c|c|c|c|c}
          & model & Ising & I+$\Delta^z_i$ & XXZ & XXZ+$\Delta^z_i$ & XXZ+$\vec\Delta_i$\\
    pulse & sequence & & & & &  \\ \hline 
    $Q_L$, & \textbf{4}  & 5  & 2  & 1 & 1 & 0 \\
    all $K=2$         & \textbf{8}  & 6  & 3  & 2 & 2 & 0 \\ 
    pulses      & \textbf{16} & 2  & 2  & 1 & 1 & 1 \\
    ($\upsilon=\alpha=0$)      & \textbf{32} & 3  & 3  & 2 & 2 & 2 \\ \hline 
    $S_L$, Herm [\onlinecite{warren-herm}]& \textbf{4}  & 3 & 1 & 1 & 1 & 0 \\    
    all $K=1$   & \textbf{8}  & 4  & 1  & 1 & 1 & 0 \\ 
    pulses         & \textbf{16} & 1  & 1  & 1 & 1 & 1 \\
    ($\upsilon=0$) & \textbf{32} & 1  & 1  & 1  & 1  & 1  \\ \hline 
          Gauss [\onlinecite{bauer-gauss}]
          & \textbf{4}  & 1 & 0 & 0 & 0 & 0 \\            
          & \textbf{8}  & 2 & 1 & 1 & 1 & 0 \\ 
          & \textbf{16} & 0 & 0 & 0 & 0 & 0 \\
          & \textbf{32} & 1 & 1 & 1 & 1 & 1 
  \end{tabular}
  \caption{Order $K$ for several decoupling sequences used  with different  
    pulse shapes upon different spin chains with nearest-neighbor and local couplings.
    Order $K$ means that the first non-zero term in the average 
    Hamiltonian~(\ref{eq:avham-expansion}) is $\bar H^{(K)}$, so that for
    small enough $\tau$ the 
    mismatch in the unitary evolution operator after $n$ decoupling cycles
    (evolution time $t=n\tau_c$) scales as 
    $\|U-\openone\|\propto t \tau_c^{K}$, and the
    corresponding infidelity $1-F\propto t^2
    \tau_\mathrm{cycle}^{2K}$.  Sequence \textbf{8} a
    sequence of $8$ pulses 
    applied intermittently on odd or even sublattices,  see
    text for actual definitions.} 
  \label{tab:decoup-order}
\end{table}

\subsection{Error scaling}
\begin{figure}[htbp]
  \centering
    \epsfxsize=0.75\columnwidth
(a)  \epsfbox{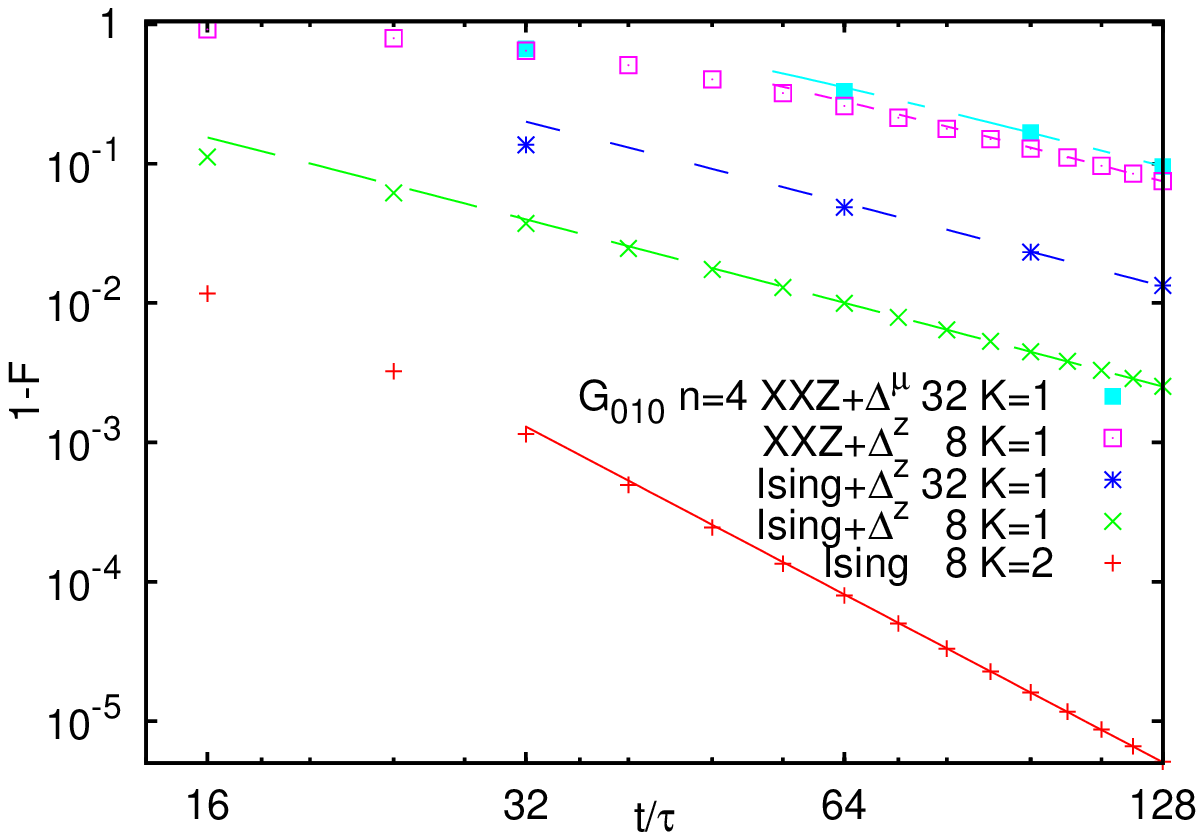} \\
    \epsfxsize=0.75\columnwidth
(b)  \epsfbox{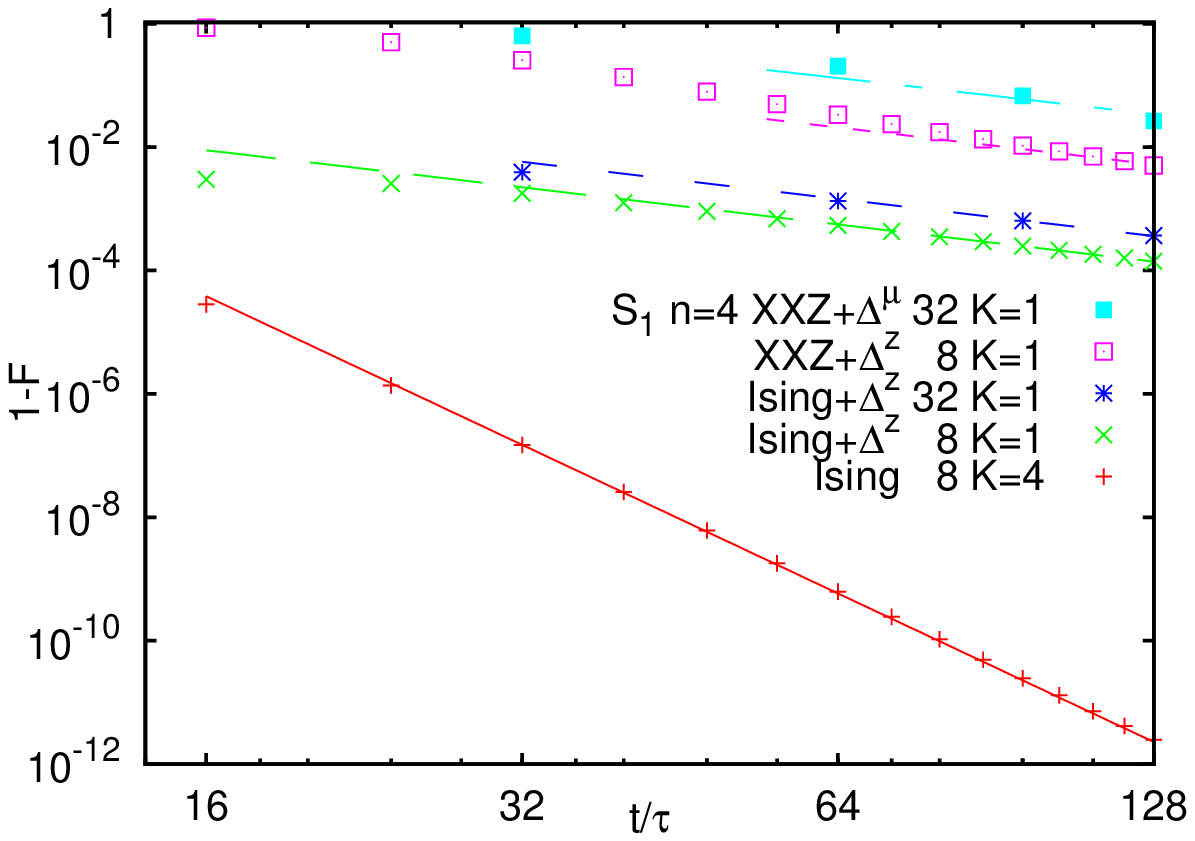} \\
    \epsfxsize=0.75\columnwidth
(c)  \epsfbox{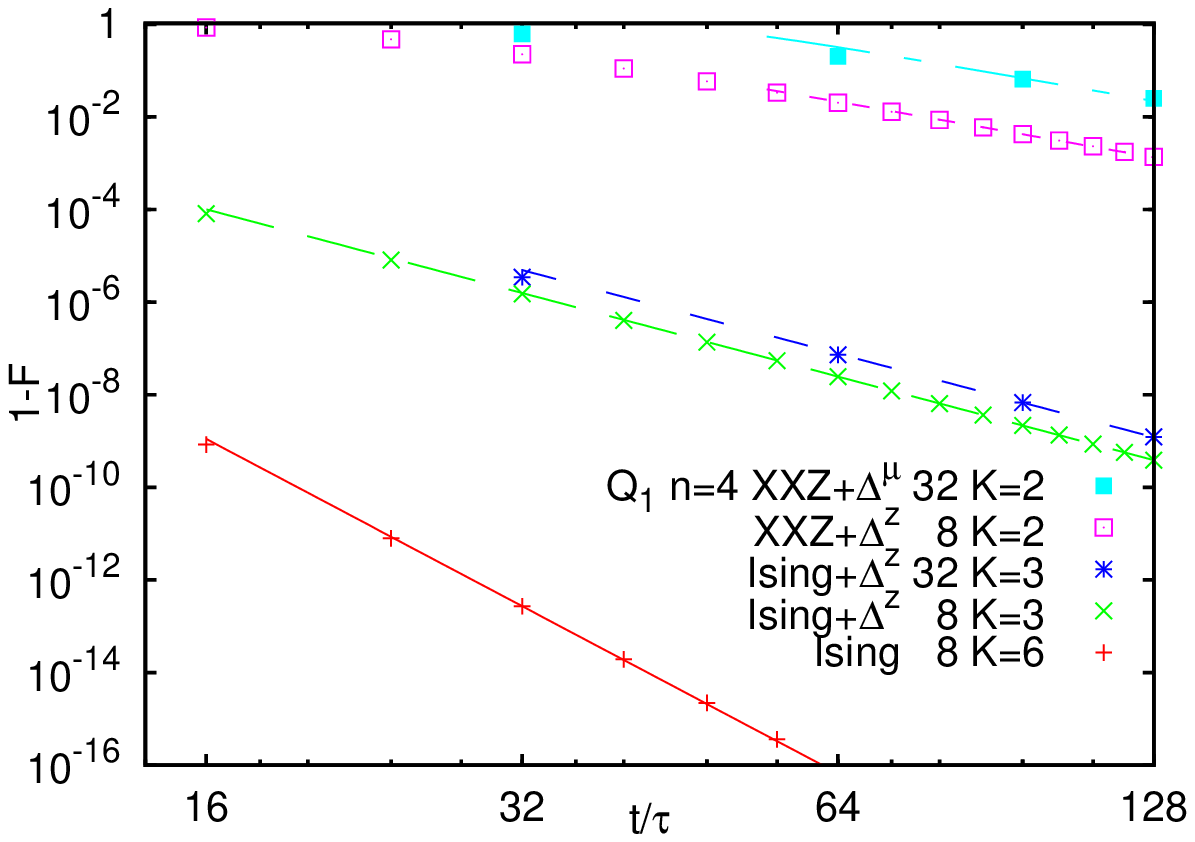} \\
\caption{(color online) Illustration of decoupling accuracy with sequences
  \textbf{8} and \textbf{32} for chains of $n=4$ qubits with different
  couplings as indicated on the plots.  The plots show
  average infidelity [see Eq.~(\protect\ref{eq:average-infidelity-delta})]
  computed at fixed time $t$ as the pulse duration $\tau$ was reduced to
  accommodate a different number of sequences. 
  The values of model parameters were
  randomly chosen and remained the same for all simulations.  
  Symbols are the data points,
  lines are the single-parameter fits of the mismatch $\delta$ [see
  Eq.~(\ref{eq:mismatch})] to $\delta=b \tau^{K}$, where the values of $K$
  indicated on the plots correspond to those in Tab.~\ref{tab:decoup-order}.
  (a) Gaussian pulses; (b) 1st-order pulses $S_1$; (c) 2nd-order pulses $Q_1$.
}
  \label{fig:samples}
\end{figure}

We illustrate the predicted power laws in Fig.~\ref{fig:samples}, where the
average infidelity (\ref{eq:average-infidelity-delta}) is computed for
different ratios of $t/\tau$, where $t$ is the fixed evolution time and the
pulse duration $\tau$ was reduced to accommodate a different number of
decoupling cycles.  The simulation is done for chains of $n=4$ qubits with
randomly chosen but fixed parameters corresponding to different chain models
as indicated.  The steepest lines correspond to largest order $K$ of the
sequence decoupling order. For symmetric sequence \textbf{8} with Ising chain,
$K=2$ for Gaussian pulses, Fig.~\ref{fig:samples}(a), $K=4$ for 1st-order
pulses, Fig.~\ref{fig:samples}(b), and $K=6$ for 2nd-order pulses,
Fig.~\ref{fig:samples}(c).  The corresponding infidelities for fixed evolution
time scale as $\propto(J_z\tau)^{4}$, $\propto(J_z\tau)^8$, and $\propto(J_z\tau)^{12}$.
Larger values of $K$ can improve accuracy by orders of magnitude, or, at fixed
required fidelity, substantially reduce the number of decoupling cycles.

We saw that with order-$K$ decoupling in multi-qubit systems with local
couplings, the decoupling error operators can be represented as connected
clusters of up to $K+1$ bonds.  For a linear chain, these involve up to $K+2$
qubits, and the number of such operators scales linearly with the total number
$n$ of qubits, as long as $n>K+2$.  In an $n$-qubit system, each of such
operators can be written as an outer product of the cluster contribution, and
the identity operators for the remaining qubits.  As a result, the square of
the Frobenius norm of the error operators scales linearly with the size of the
Hilbert space, that is, \emph{exponentially} with the number of qubits.
However, this exponential scaling is suppressed when we compute the infidelity
[see Eq.~(\ref{eq:average-infidelity-delta})], so that the infidelity scales
only \emph{linearly} with the number of clusters, that is, linearly with the
number of qubits.  The same scaling with the system size is expected in higher
dimensional arrangements of qubits (planar, 3D).

\begin{figure}[htbp]
  \centering
  \epsfxsize=0.75\columnwidth
  \epsfbox{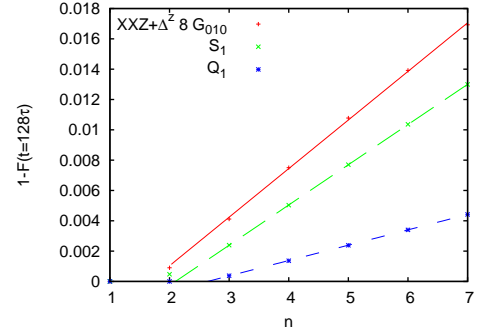} \\
  \caption{(color online) Scaling of the infidelity $1-F$ at $t/\tau=128$ with
    the chain length $n$ for a particular realization of an XXZ chain with
    on-site disorder $\Delta^z_i$, decoupling sequence \textbf{8}, pulse
    shapes as indicated.  (Data for the pulse $G_{010}$ divided by 10 to
    fit with the other data.) While the unitary matrix mismatch $\delta^2$ [see
    Eq.~(\ref{eq:mismatch})] grows exponentially with the chain length $n$,
    $\delta^2\propto 2^n$, the leading-order contribution to the corresponding
    infidelity $(1-F)$ represents the probability of error in one of the
    clusters, and it scales only linearly, as also seen in the plots.}
  \label{fig:size}
\end{figure}

We illustrate the scaling of decoupling errors with the qubit number $n$ in
Fig.~\ref{fig:size}.  The plots show the scaling of the average infidelity at
the end of the interval in Fig.~\ref{fig:samples} and other data with the
chain length $n$.  

\subsection{Composite pulses}
Composite pulses are, in fact, pulse sequences designed to replace a single
pulse and specially designed to compensate for some particular systematic
errors, including off-resonance application, pulse amplitude, and pulse phase
errors\cite{Mehring-1990,wimperis-1994,Cummins-2003,hugh:012327,%
  PhysRevA.70.052318,Hedin-2000,Blatt-2003,Jones-2003,Mignon-2004}.

The off-resonance errors appear when the carrier frequency of the applied
pulse is off the transition frequency between the $|0\rangle$ and $|1\rangle$
state of a qubit.  In the rotating reference frame this is equivalent to a
non-zero chemical shift Hamiltonian~(\ref{eq:chemical-shift}), with $\Delta$
equal to the corresponding frequency bias.  We note that our 1st and
2nd-order self-refocusing pulse shapes already offer a degree of stability
against such errors.  

For this reason we concentrate on the pulse \emph{amplitude errors}, where the
correct pulse shape is applied with the wrong amplitude, producing an
incorrect rotation angle $\tilde\phi_0\neq \phi_0$.  Note that no
one-dimensional pulse shaping can compensate for this kind of errors, since
the  modified rotation angle is simply proportional to the pulse amplitude,
$\tilde \phi_0=(1+f) \phi_0$.

On the other hand, one can expect that the pulse amplitude offset $f$ remains
the same for all the pulses applied at a particular frequency.  This
uniformity is utilized in several composite pulses designed so that the net
rotation would be insensitive to such uniform errors.

\subsubsection{SCROFULOUS}
\begin{figure}[htbp]
   \centering 
   (a)\epsfxsize=0.75\columnwidth
   \epsfbox{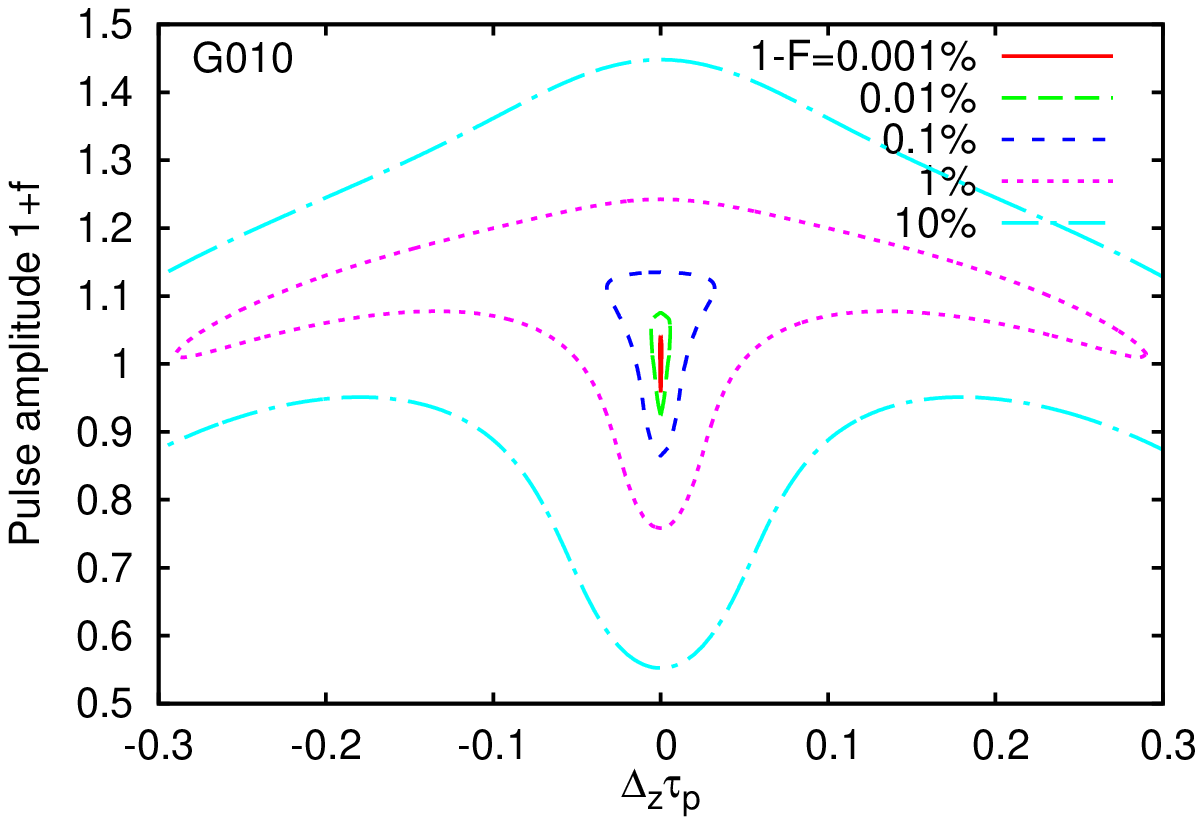}\\
   (b)\epsfxsize=0.75\columnwidth
   \epsfbox{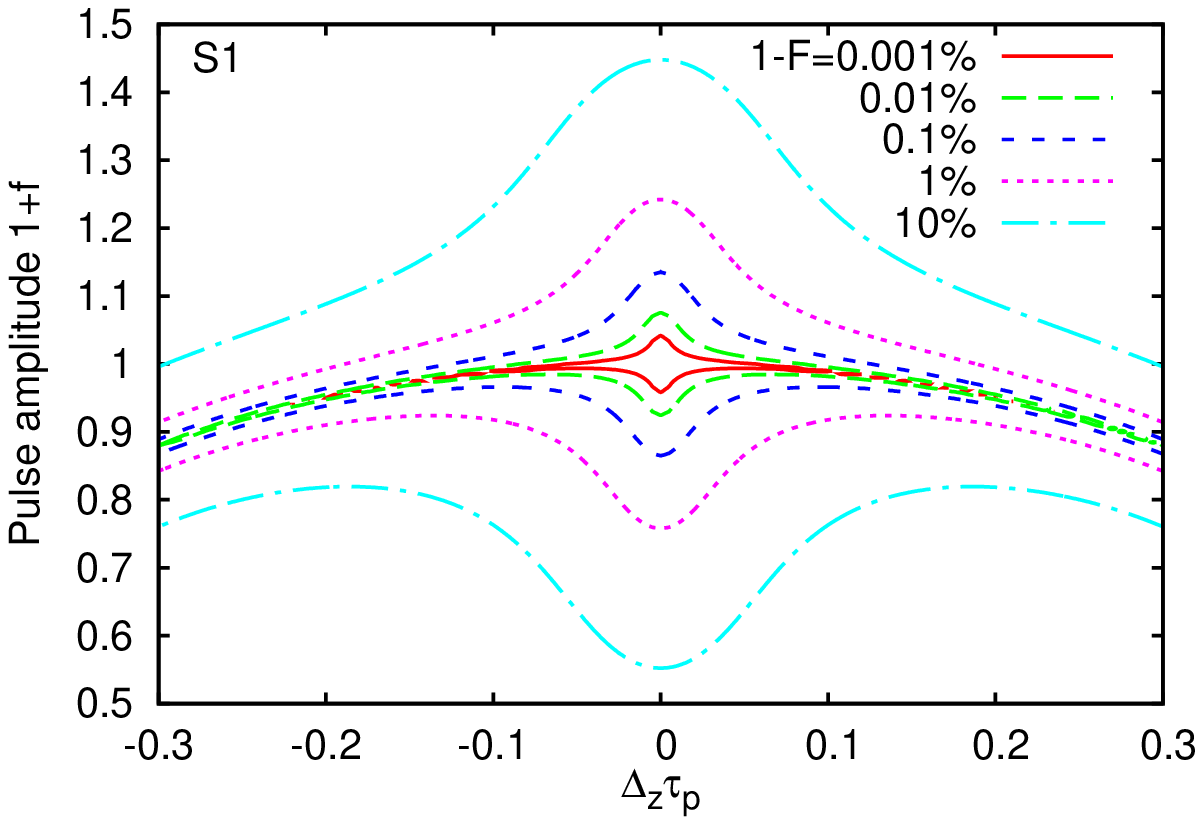}
   (c)\epsfxsize=0.75\columnwidth
   \epsfbox{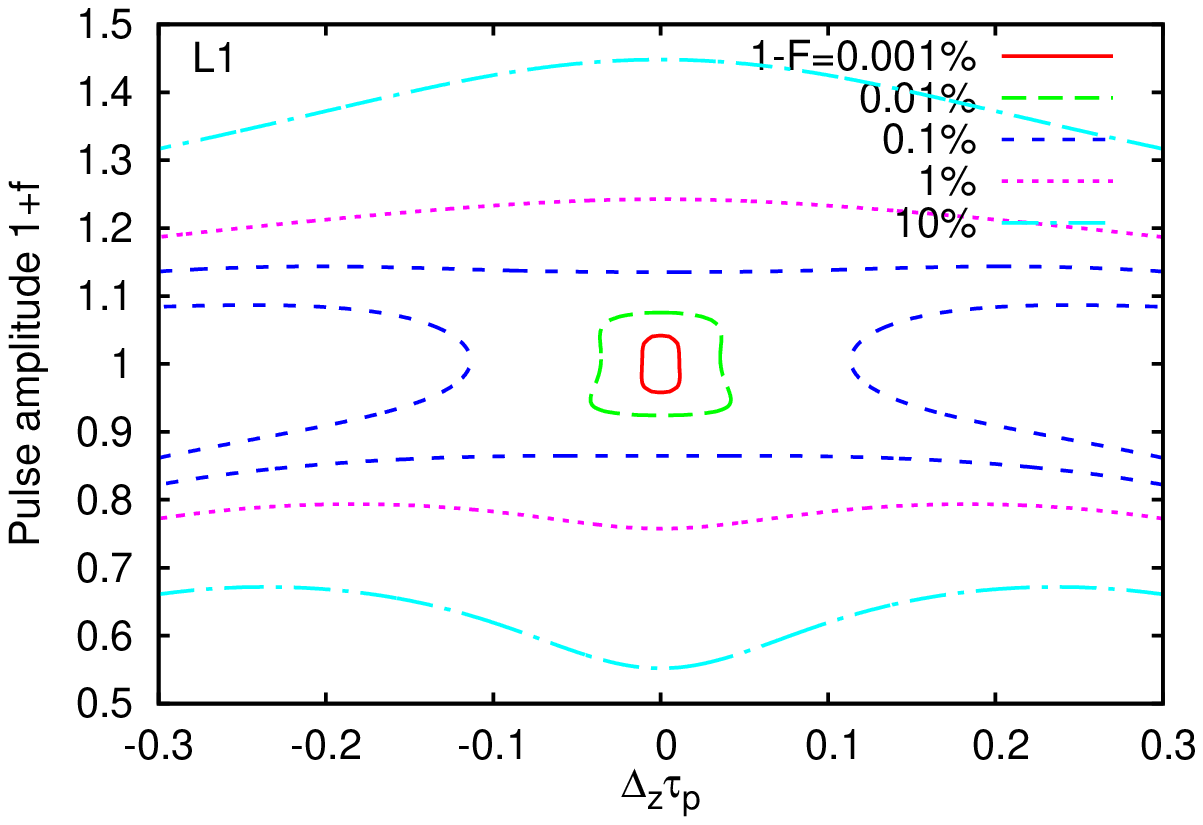}
   (c)\epsfxsize=0.75\columnwidth
   \epsfbox{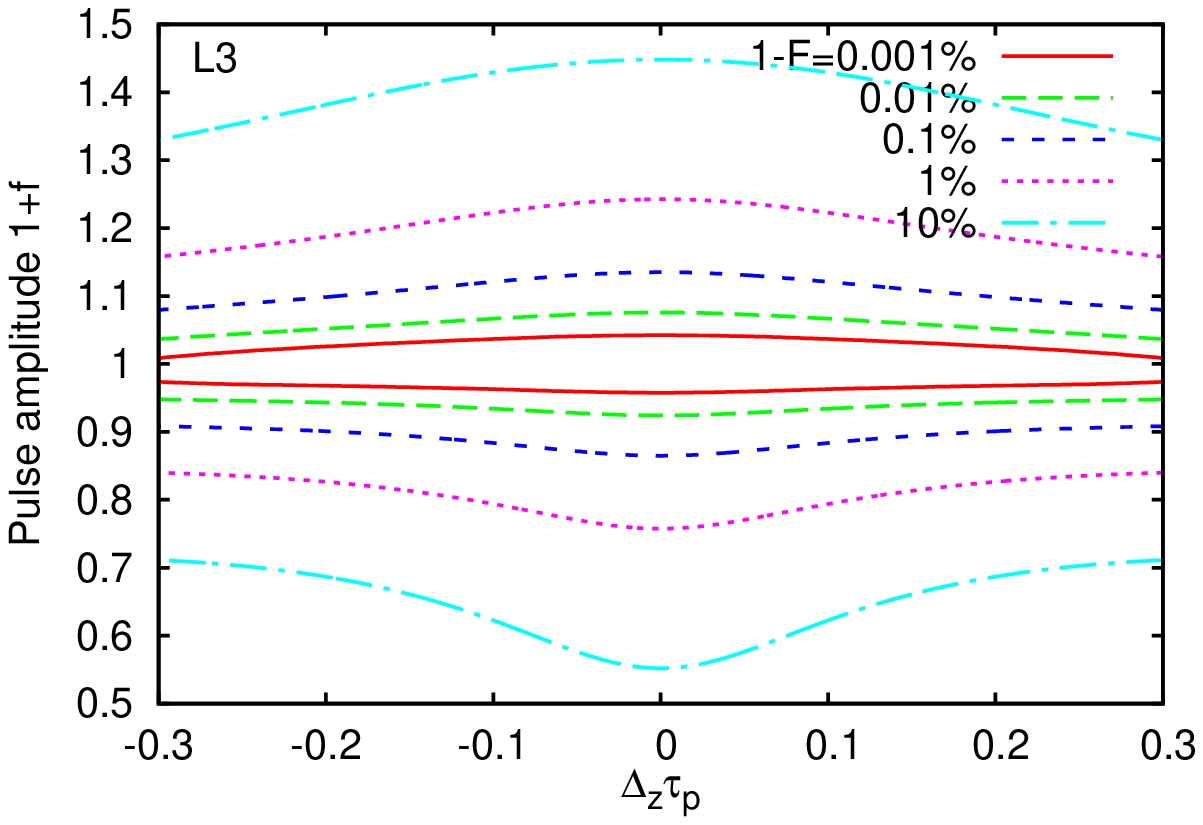}
   \caption{(color online) Contour plots of the average fidelity for the
     composite pulse SCROFULOUS $\pi_{60}\pi_{300}\pi_{60}$ with (a) Gaussian
     pulses $G_{010}$; (b) 1-st order self-refocusing pulses $S_1$ (the plots
     for pulses $Q_1$ look similarly but symmetric with respect to horizontal
     axis); (c) 1st-order pulse with amplitude correction
     $\upsilon=\upsilon'=\upsilon''=0$; (d) 2nd-order pulse with amplitude
     correction, $\upsilon=\upsilon'=\upsilon''=\alpha=0$.  The axes are the
     relative frequency mismatch $\tau \Delta$ and the relative pulse
     amplitude $(1+f)$, see text.}
   \label{fig:scr}
 \end{figure}

The three-pulse sequence {\tt SCROFULOUS}\cite{Cummins-2003} is based on the
sequence originally proposed by Tycko\cite{tycko-1983,Tycko-1984}.
Particularly, an improved $\pi$ pulse is obtained by applying three $\pi$
pulses, at $60^\circ$, $300^\circ$, and again at $60^\circ$, or just
$\pi_{60}\pi_{300}\pi_{60}$.  In the case of ideal $\delta$-pulses, the
resulting pulse compensates for pulse amplitude errors to linear order.  With
finite-width shaped pulses, an additional error is generated due to the
presence of the system Hamiltonian.  In particular, for the chemical-shift
Hamiltonian~(\ref{eq:chemical-shift}), the expansion of a unitary operator
applied along $x$ axis has the form [see Eqs.~(\ref{eq:zeroth-order-ham-cs}),
(\ref{eq:ham-chem-1})]
\begin{eqnarray}
  \label{eq:unitary-x-cs} \nonumber 
  U_x& =& \cos
   \frac{\phi _0}{2} -i
   \sigma _x \sin \frac{\phi
   _0}{2} \\ \nonumber 
& & -i { \tau\Delta \upsilon\over 2}  \left(\cos
   \phi _0 \sigma _z-\sigma _y \sin
   \phi _0\right) \\ \nonumber 
& & 
+ {\tau^2\Delta^2\upsilon^2\over 8}  \left(i \sigma _x \sin
   \frac{\phi _0}{2}-\cos
   \frac{\phi _0}{2} \right)\\
& &    +{\tau^2\Delta
   ^2\alpha\over 4}  \left(i\sigma _x \cos
   \frac{\phi _0}{2} +
   \sin \frac{\phi
   _0}{2}\right)+{\cal O}(\tau)^3.\quad 
\end{eqnarray}
Combining the corresponding expressions appropriately rotated in the $x$-$y$
plane, and expanding the result to quadratic power in the relative amplitude
offset $f$, we obtain for the composite pulse $\pi_{60}\pi_{300}\pi_{60}$,
\begin{equation}
  \label{eq:scr-expansion}
U_\mathrm{SCR}=-i \sigma _x
+i {\tau  \Delta 
  \tilde \upsilon}  \sigma _z  -{i}{ \sqrt{3}   \pi^2 f^2\over 8}\sigma_y
+\ldots, 
\end{equation}
where 
$\tilde\upsilon\equiv \tilde\upsilon(f)=\upsilon+\upsilon' f+{\cal O}(f^2)$ is
the parameter $\upsilon$ [Eq.~(\ref{eq:upsilon-defined})] but for the pulse
with rescaled amplitude, and the further terms are of order $f\tilde\upsilon
\tau\Delta$, $\tilde\upsilon^2 \tau^2\Delta^2$, $\tilde\alpha f \tau^2
\Delta^2$.  Clearly, with Gaussian or other pulse shape such that
$\upsilon\neq0$, the error is linear in $\tau \Delta$ and quadratic in the
amplitude shift $f$ (although generally there will also be a cross-term
$\propto f\tau\Delta$).  This situation is illustrated in
Fig.~\ref{fig:scr}(a), where the average infidelity is plotted for the
SCROFULOUS sequence with pulses $G_{010}$ on the plane $\Delta\tau$--$f$.  The
region for $1-F=10^{-5}$ is a narrow vertical line which corresponds to great
sensitivity to frequency shift.  With 1st-order self-refocusing pulses such
that $\upsilon=0$, $\tilde\upsilon(f)\propto f$, the error is dominated by the
term $\propto f\tau\Delta$.  The corresponding region corresponds to a
diamond-like shape in the center of Fig.~\ref{fig:scr}(b).  We have also
generated self-refocusing pulse shapes such that both $\upsilon=0$ and
$\upsilon'=0$.  Then, by symmetry, $\upsilon''=0$, and generically
$\tilde\upsilon\propto f^3$.  Then, for 1st-order pulses, $\alpha\neq0$, the
errors are dominated by the term omitted in Eq.~(\ref{eq:scr-expansion}); they
scale as ${\cal O}(f^2)$, ${\cal O}(\upsilon'''\Delta f^3)$, ${\cal
  O}(\alpha\Delta^2)$, while for second-order pulses the last two terms become
${\cal O}(\Delta^3)$, ${\cal O}(\alpha' f^2 \Delta^2)$.  Plots for such shapes
are shown in Figs.~\ref{fig:scr}(c) and~\ref{fig:scr}(d) respectively; the
result of improved pulse stability is a much wider region of high fidelity.  

\subsubsection{BB$_1$ and related pulses}
A longer but more accurate composite pulse known as BB$_1$ was
originally proposed by Wimperis\cite{wimperis-1994}.  For target angle
$\theta=\pi$, the pulse can be written as BB$_1^{(W)}=\pi_0
\pi_\phi(2\pi)_{3\phi}\pi_\phi$, where
$\phi=-\cos^{-1}(-1/4)\approx104.5^\circ$.  For ideal $\delta$-pulses,
this cancels errors of both 1st and 2nd order in the relative pulse
amplitude bias $f$.  A related symmetrized sequence
BB$_1^{(CLJ)}=(\pi/2)_0 \pi_\phi(2\pi)_{3\phi}\pi_\phi(\pi/2)_0$ was
proposed in Ref.~\onlinecite{Cummins-2003} [see also
  Ref.~\onlinecite{hugh:012327}]; because of the symmetry it leads to
some additional error cancellation at higher order.  With shaped
pulses, we have also analyzed variants of these sequences with the
$2\pi$ pulses replaced by two $\pi$ pulses, BB$_1^{(W')}=\pi_0
\pi_\phi(\pi)_{3\phi}(\pi)_{3\phi}\pi_\phi$ and
BB$_1^{(CLJ')}=(\pi/2)_0
\pi_\phi(\pi)_{3\phi}(\pi)_{3\phi}\pi_\phi(\pi/2)_0$.

Computing the products of versions of Eq.~(\ref{eq:unitary-x-cs})
appropriately rotated in the $x$-$y$ plane, for on-resonance application of
any version of the BB$_1$ sequence, 
\begin{equation}
U_{{\rm BB}_1}^{(\Delta=0)}=-i\sigma_x -{f^3\pi^3\over 64}(5-i
15^{1/2}\sigma_z)+{\cal O}(f^4). \label{eq:bb1-amplitude}
\end{equation}
We note that to achieve the level of infidelity of, say, $1-F=10^{-4}$, the
frequency mismatch should satisfy $|f|< 0.136$, compared with $|f|< 0.090$ for
the SCROFULOUS sequence.  Thus, even though order of the BB$_1$ family of
composite pulses is higher (and thus, for small $|f|$ their performance is
much better asymptotically), at this level of infidelity their performance is
comparable.

\begin{figure}[htbp]
   \centering 
   (a)\epsfxsize=0.75\columnwidth
   \epsfbox{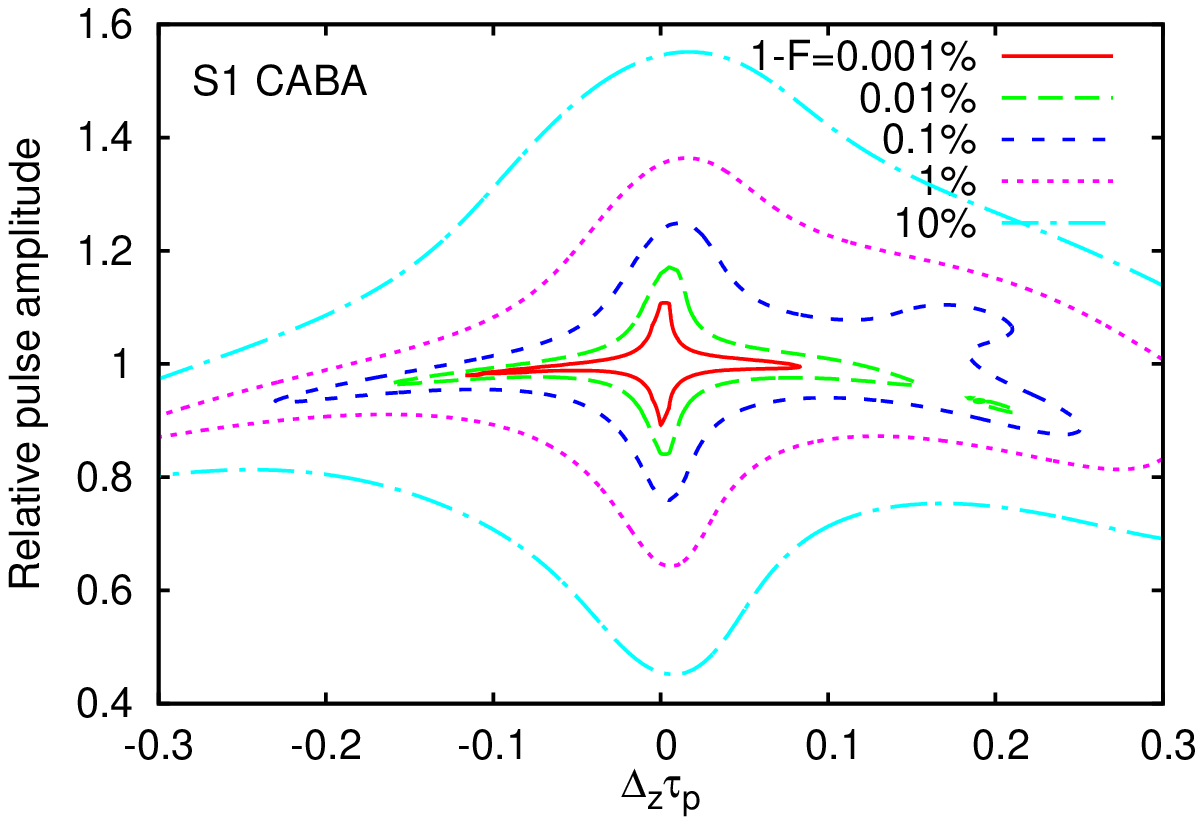}
   (b)\epsfxsize=0.75\columnwidth
   \epsfbox{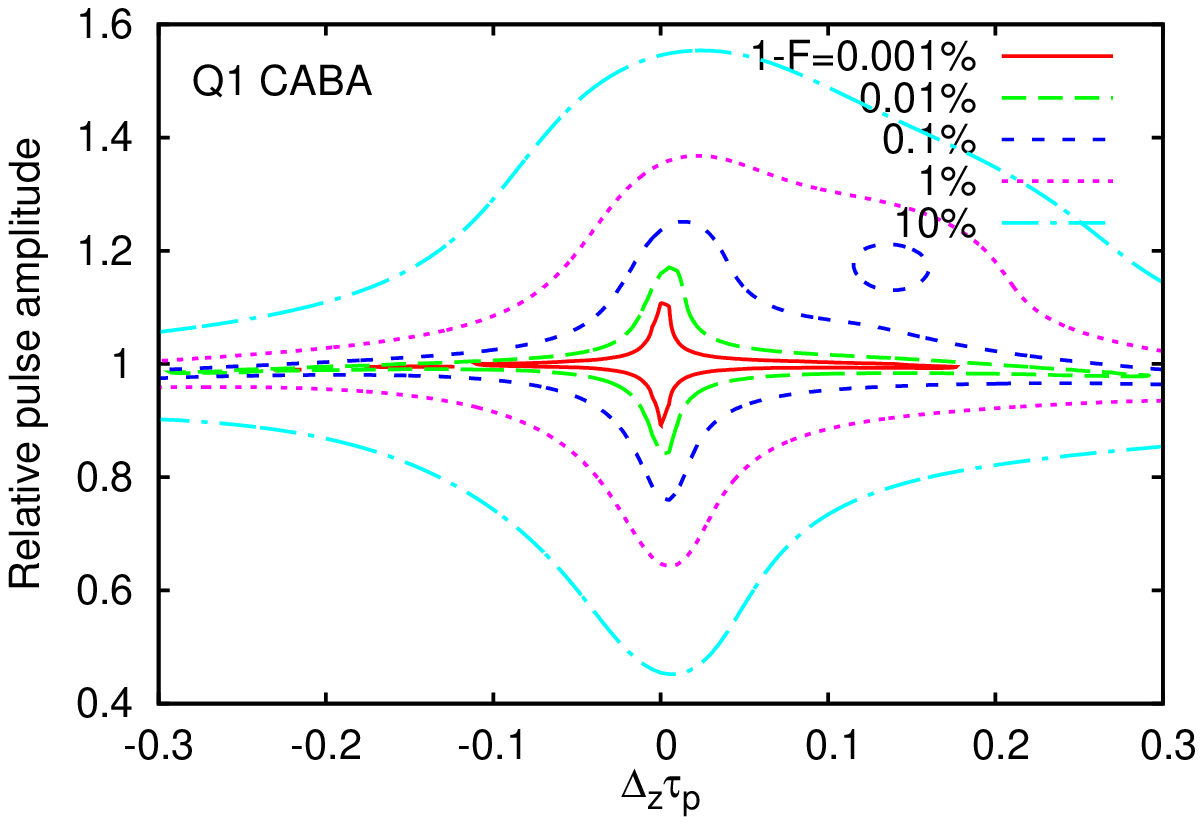}
   (c)\epsfxsize=0.75\columnwidth
   \epsfbox{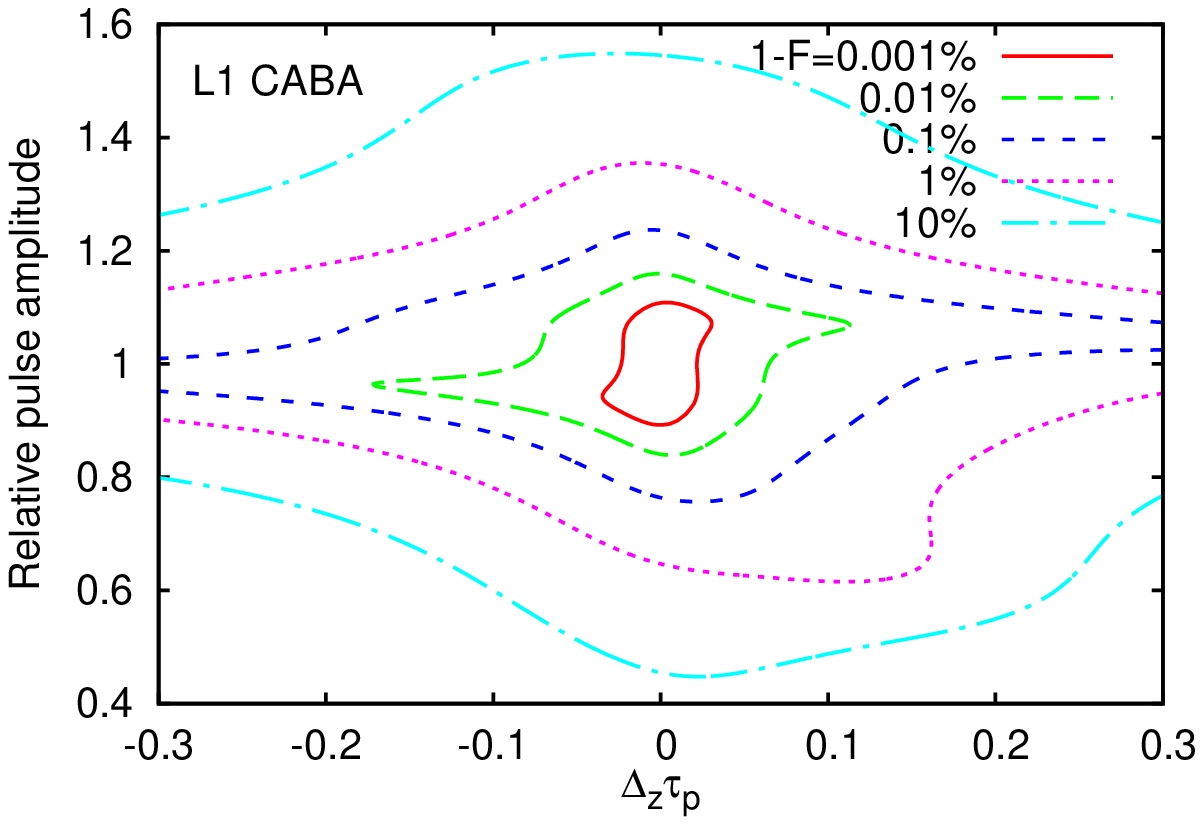}
   (d)\epsfxsize=0.75\columnwidth
   \epsfbox{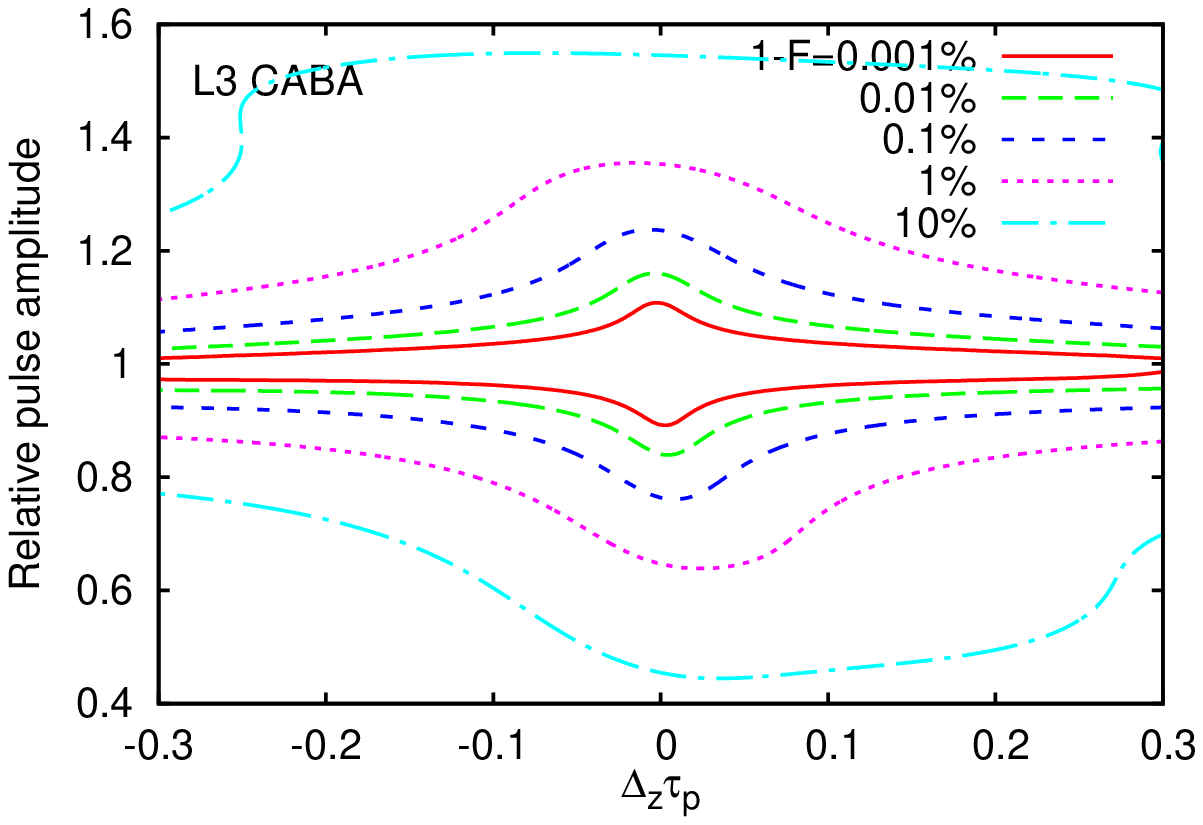}
   \caption{(color online) Contour plots of the average infidelity $1-F$ for
     the composite pulse BB$_1^{(W)}$ [$\pi_0\pi_\phi(2\pi)_{3\phi}\pi_\phi$]
     with (a) 1st-order self-refocusing pulses $S_1$; (b) 2nd-order pulses
     $Q_1$; (c) 1st-order pulses with amplitude correction; (d) 2nd-order
     pulses with amplitude correction.  The axes are the relative frequency
     mismatch $\tau \Delta$ and the relative pulse amplitude $(1+f)$, see
     text.}
   \label{fig:bb1-W}
 \end{figure}

\begin{figure}[htbp]
   \centering 
   (a)\epsfxsize=0.75\columnwidth
   \epsfbox{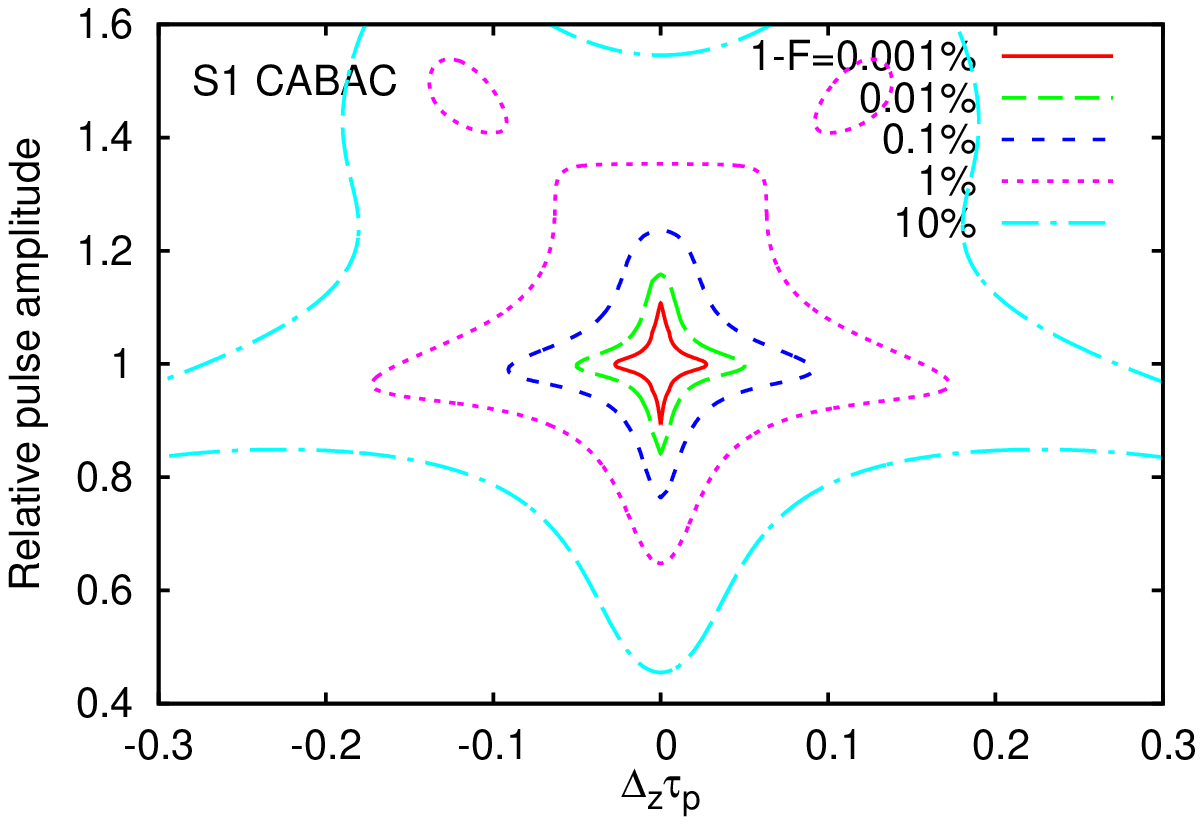}
   (b)\epsfxsize=0.75\columnwidth
   \epsfbox{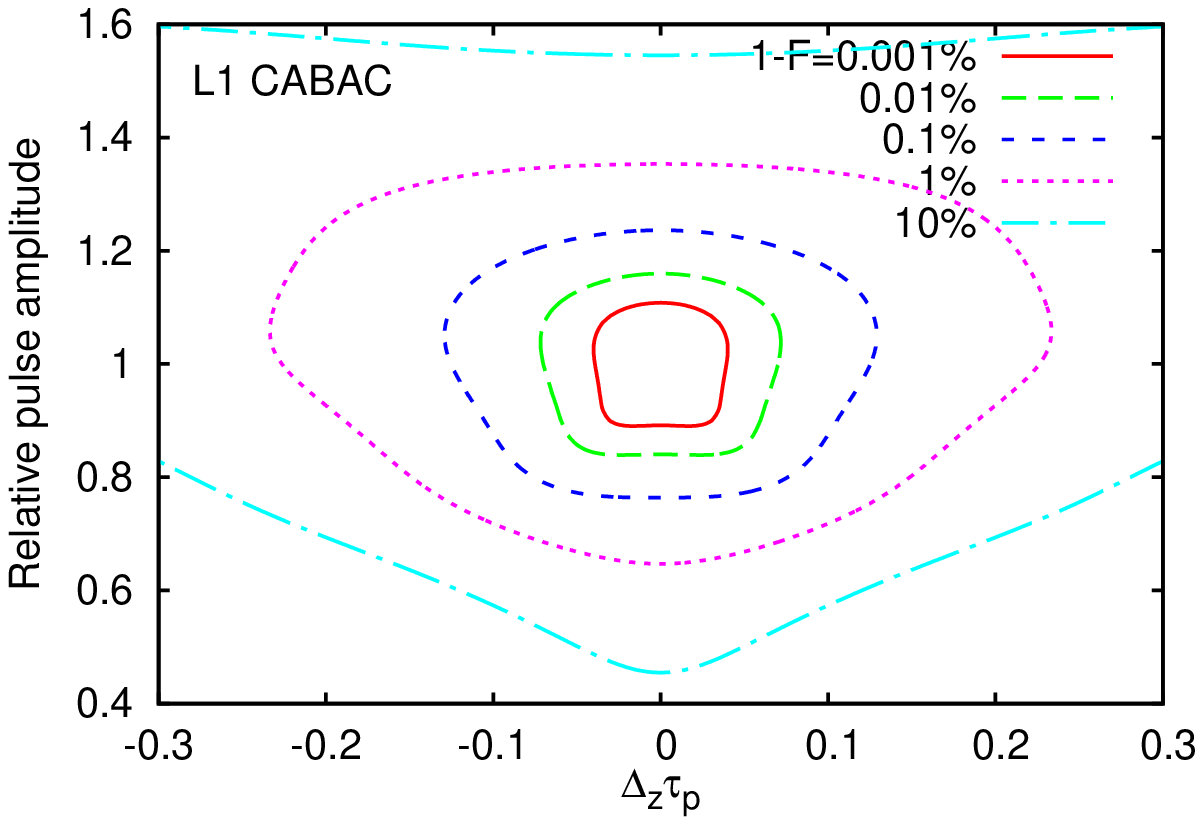}
   (c)\epsfxsize=0.75\columnwidth
   \epsfbox{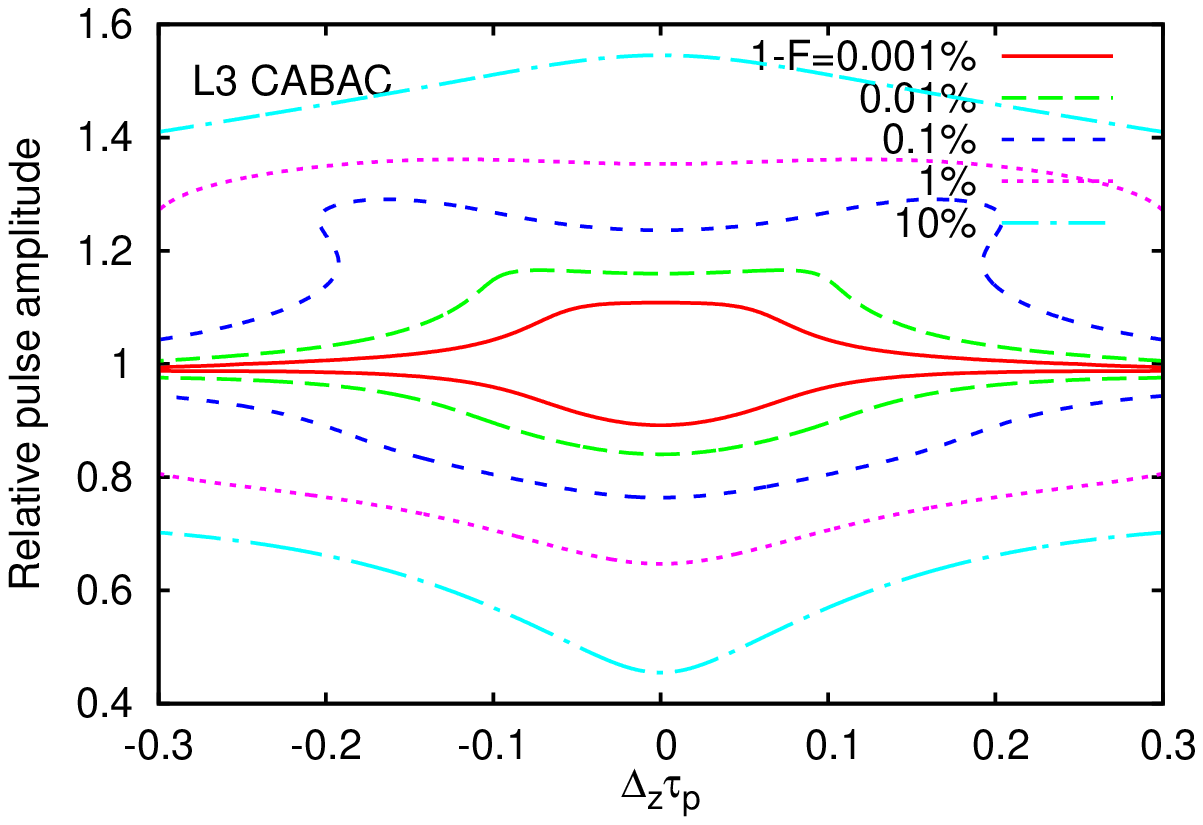}
   \caption{(color online) As in Fig.~\ref{fig:bb1-W} but for the
     symmetrized sequence BB$_1^{(CLJ)}$
     [$(\pi/2)_0\pi_\phi(2\pi)_{3\phi}\pi_\phi(\pi/2)_0$].  The pulse shapes
     are (a) 1st-order pulses $S_1$ (the fidelity for the 2nd-order pulses
     $Q_1$ is similar); (b) 1st-order pulses with amplitude correction; (c)
     2nd-order pulses with amplitude correction.  Note how regular are the
     shapes of high-fidelity regions.}
   \label{fig:bb1-CLJ}
 \end{figure}

We now turn to off-resonance correction terms which differ between
implementations of the BB$_1$ sequence.  In particular, with generic pulses
such that $\upsilon\neq0$, already at $f=0$, all of these sequences acquire
linear corrections scaling with $\tau\Delta $.  For example, the expansion of
the sequence BB$_1^{(W)}$ at $f=0$ can be written as
\begin{eqnarray}
  \label{eq:bb1-off-resonance}
  \nonumber 
  \lefteqn{  U_{{\rm BB}_1}^{(W,f=0)}=-i \sigma _x
    +i\frac{\tau\Delta}{2}  \left(\sigma _z
       \upsilon _1-\sigma _y  \upsilon _2\right)} & & \\
  & & +\frac{ \tau^2\Delta^2}{16}  \left(2 i \sigma _x  \upsilon^2+6  \upsilon_2
     \upsilon_1+ \alpha+i 
    \sqrt{15} \sigma _z \left( \alpha-2  \upsilon _1  \upsilon
      _2\right)\right) \nonumber \\ 
  & & + {\cal O}(\Delta^3\tau^3) ,
\end{eqnarray}
where $ \alpha\equiv 2  \alpha _1- \alpha _2$,
$ \upsilon^2\equiv \upsilon _1^2+ \upsilon_2^2$, and the
parameters $ \upsilon_1$, $ \alpha_1$ and $ \upsilon_2$,
$ \alpha_2$ correspond to the $\pi$ and $2\pi$ pulses respectively.
Notice that the second-order coefficients $\alpha_i$ enter only in the
combination $2\alpha_1-\alpha_2$.  Not surprisingly, if we replace the
$2\pi$-pulse with two $\pi$ pulses, the coefficients $\alpha$ cancel out,
\begin{eqnarray}
  \label{eq:bb1-off-resonance-prime}
  U_{{\rm BB}_1}^{(W',f=0)}& =& -i \sigma _x
  +i\frac{\tau\Delta\upsilon _1}{2} 
  \sigma _z 
  +i\frac{\tau ^2\Delta ^2\upsilon_1^2 }{8}   \sigma _x;
\end{eqnarray}
thus second-order accuracy can be obtained already with 1st order pulses,
$\upsilon_1=0$.  Now, when both the amplitude and the resonant frequency bias
are present, $f\neq0$ and $\Delta\neq0$, there is an additional source of
error due to dependence of the pulse parameters on the amplitude,
$\upsilon,\alpha\to \tilde\upsilon(f), \tilde\alpha(f)$.  For regular
self-refocusing pulses, $\tilde\upsilon(f)=\upsilon' f+{\cal O}(f^2)$, and the
1st order terms in
Eqs.~(\ref{eq:bb1-off-resonance}),~(\ref{eq:bb1-off-resonance-prime}) and
their analogs for the other variants of the BB$_1$ pulse dominate the error
$\propto f\tau \Delta $.  Respectively, the average infidelity scales as $\propto(f
\tau\Delta)^2$, resulting in characteristic diamond-like shape on the contour
plots of infidelity, see Figs.~\ref{fig:bb1-W}(a,b), \ref{fig:bb1-CLJ}(a).

\begin{figure}[htbp]
   \centering 
   (a)\epsfxsize=0.75\columnwidth
   \epsfbox{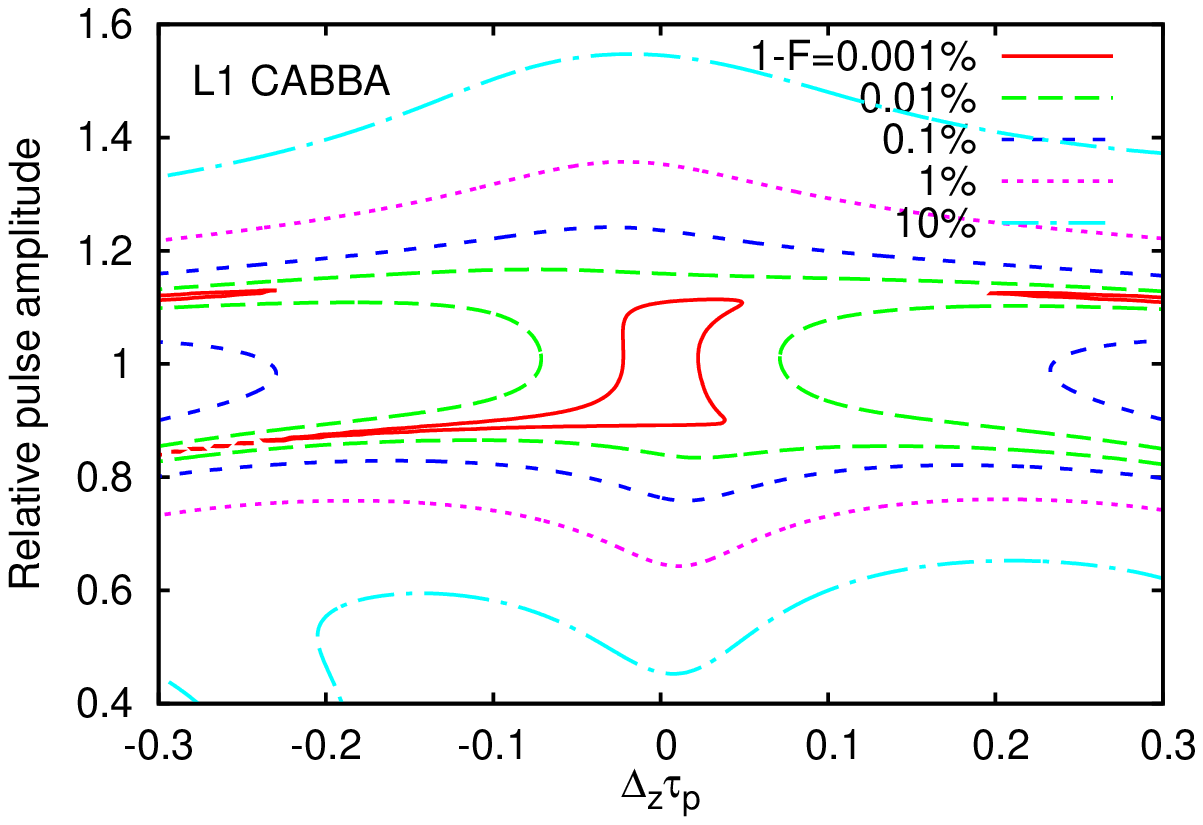}
   (b)\epsfxsize=0.75\columnwidth
   \epsfbox{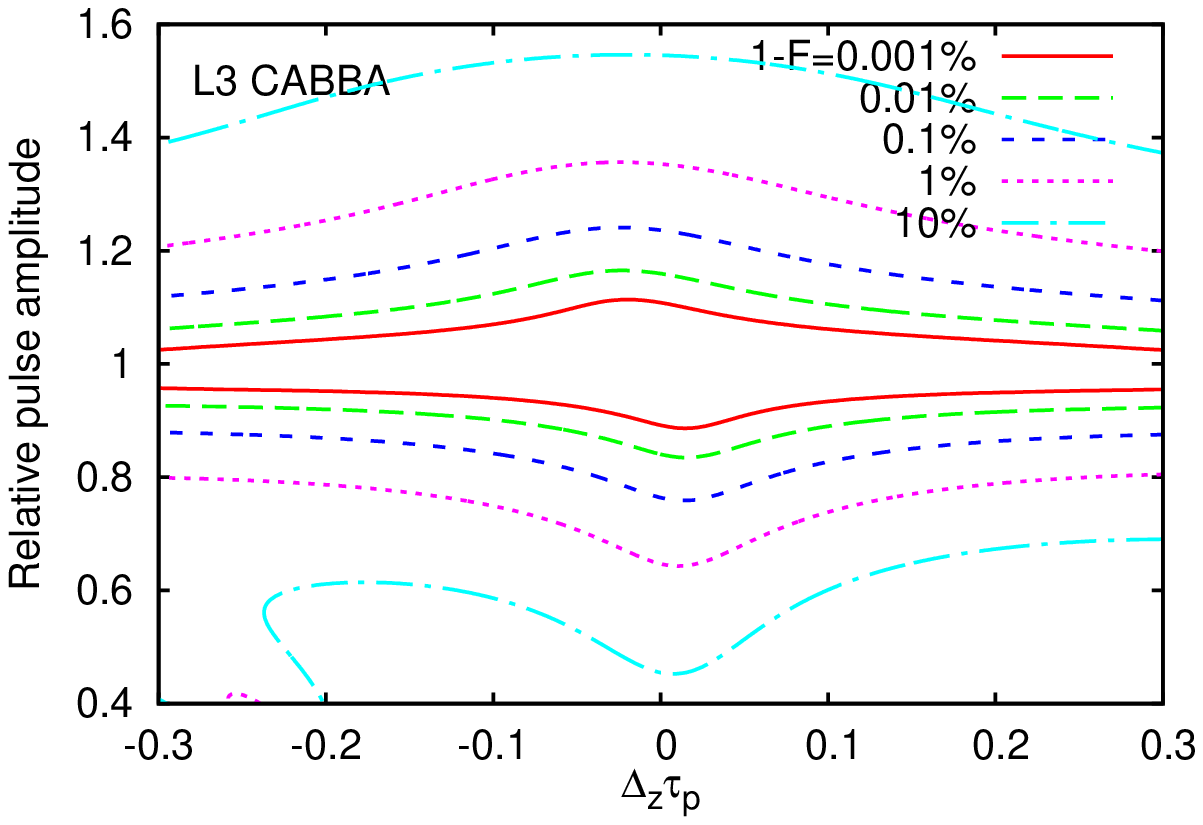}
   \caption{(color online) As in Fig.~\ref{fig:bb1-W} but for the sequence
     BB$_1^{W'}$ [$\pi_0\pi_\phi\pi_{3\phi}\pi_{3\phi}\pi_\phi$] using only
     $\pi$-pulses.  (a) 1st-order pulses with amplitude correction; (b)
     2nd-order pulses with amplitude correction.  The absence of the error
     terms linear in $\alpha$ [see Eqs.~(\ref{eq:bb1-off-resonance-prime}),
     (\ref{eq:bb1-W-crossterm})] produces a much wider high-fidelity region
     already with 1st-order pulses.}
   \label{fig:bb1-Wp}
 \end{figure}

With specially designed pulses such that both $\upsilon=0$ and $\upsilon'=0$,
due to pulse symmetry, also $\upsilon''=0$, so that $\upsilon(f)=\upsilon'''
f^3+{\cal O}(f^4)$.  Then, for sequences other than BB$_1^{(W')}$, with
1st-order pulses the error is dominated by the terms quadratic in $\tau\Delta$
due to coefficients $\alpha_i$ [cf.\ Eq.~(\ref{eq:bb1-off-resonance})].  As a
result, the high-fidelity regions in Figs.~\ref{fig:bb1-W}(c),
\ref{fig:bb1-CLJ}(b), and \ref{fig:bb1-Wp}(a) have much more rounded shape.
With 2nd-order pulses with amplitude correction, such that $\alpha_i=0$ but
$\alpha_i'\neq0$, the leading-order error scales as ${\cal O}(f
\tau^2\Delta^2)$, which extends the high-fidelity regions out to larger values
of $\tau \Delta$ in a characteristic ``smile'' pattern.  For the sequence
BB$_1^{(W')}$ these terms cancel out, and the leading-order error term comes
from the non-zero $\upsilon'''$, the errors scale as $\propto f^3 \tau\Delta$,
\begin{equation}
  \label{eq:bb1-W-crossterm}
  U_{{\rm BB}_1}^{(W')}=U_{{\rm BB}_1}^{(W',f=0)}+i
  {f^3\tau\Delta\,\upsilon_1''' \over 2}\sigma_z+{\cal O}(\tau^3 \Delta^3). 
\end{equation}
This error has the same order in $f$ as that in Eq.~(\ref{eq:bb1-amplitude}),
and it can compensate or increase the contribution linear in $\sigma_z$.  The
result is a somewhat skewed in the center high-fidelity region widely
stretched horizontally.

\subsection{Stability of decoupling against amplitude errors}
We now return to the problem of decoupling for a chain of qubits, but now
consider the effect of the amplitude errors.  This needs a separate study
since single-qubit errors with composite pulses have structure different from
those due to, say, finite pulse width.

\begin{figure}[htbp]
   \centering 
   (a)\epsfxsize=0.75\columnwidth
   \epsfbox{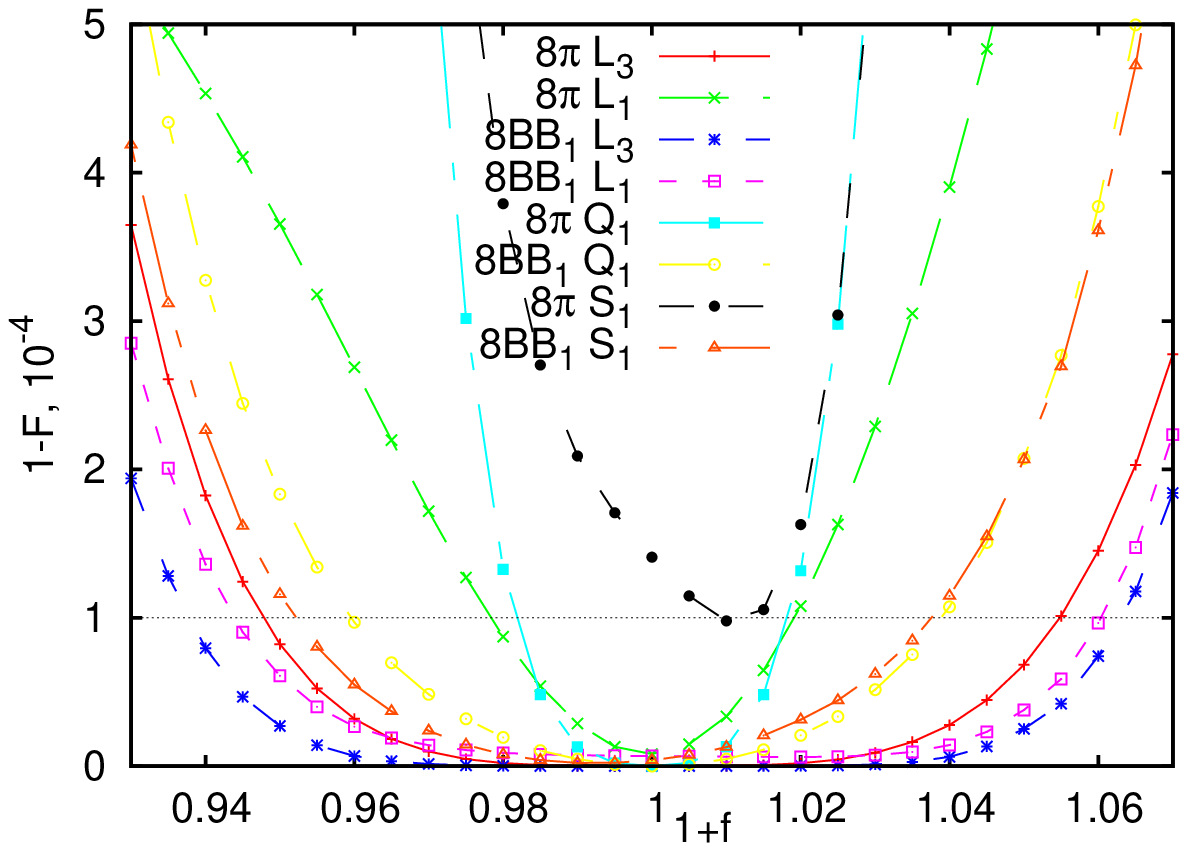}
   (b)\epsfxsize=0.75\columnwidth
   \epsfbox{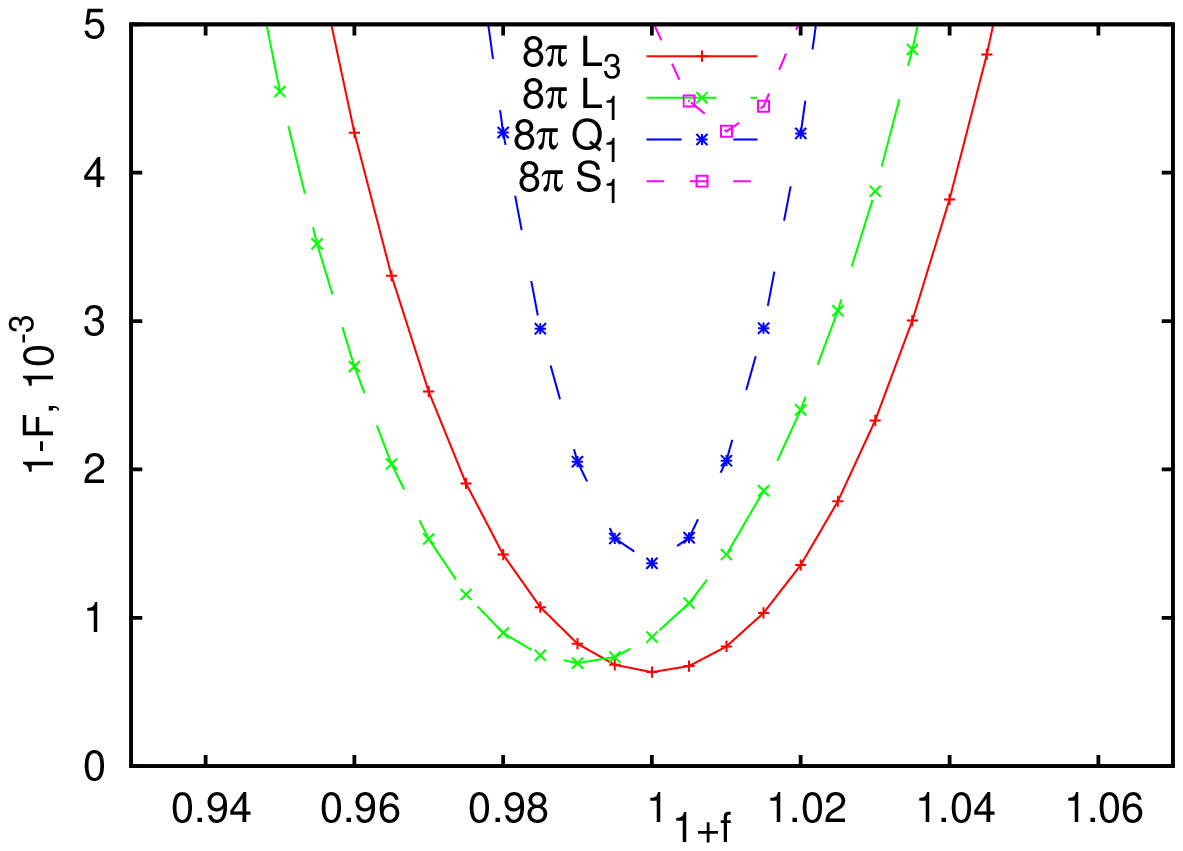}
   \caption{(color online) Decoupling errors as a function of relative pulse
     amplitude for a chain of $n=4$ qubits with different decoupling schemes
     as indicated.  (a) Ising chain in the presence of individual chemical
     shifts $\Delta^z_i$.  8-pulse sequence with BB$_1$ composite pulses
     offers the best accuracy, which remains essentially the same whether the
     sequence is used with regular 1st- or 2nd- order pulses or with the
     pulses stabilized against amplitude errors (1st-order pulses $L_1$ and
     2nd-order pulses $L_3$).  However, the pulse $L_3$ works well enough even
     with regular 8-pulse sequence. While the details of the amplitude scaling
     differ, at the level of $1-F=10^{-4}$, the 1st-order amplitude-protected
     pulses $L_1$ and regular 2nd-order pulses $Q_1$ have comparable
     accuracy.  The use of 1st-order pulses show relatively poor performance
     even on resonance. (b) XXZ chain in the presence of individual chemical
     shifts $\Delta^z_i$.  With XXZ coupling, The BB$_1$ composite pulse is no
     longer accurate, as the errors appear already in the linear order in
     $\tau J^\perp$ (not shown).  With the regular 8-pulse sequence
     \textbf{8}, the best accuracy is obtained for the pulses with amplitude
     correction.}
\label{fig:8_bb1}
 \end{figure}

In Fig.~\ref{fig:8_bb1}(a) we present the results of simulations for a
particular 4-qubit Ising chain with on-site chemical potential shifts
$\Delta^z_i$ over time interval $t=128\tau$ (exactly the same parameters as in
Figs.~\ref{fig:samples}, \ref{fig:size}).  Specifically, we plot the
infidelity $(1-F)$ in units of $10^{-4}$, as a function of the fractional
pulse amplitude $1+f$.  In the first half of the symmetric sequence
$\textbf{8}=\X_1\Y_2\bar\X_1\bar Y_2\,\bar Y_2\bar\X_1\Y_2\X_1$, we used the
original BB$_1^{(W)}$ pulse $\X\to\pi_0\pi_\phi(2\pi)_{3\phi}\pi_\phi$ for
$\pi_x$, rotated appropriately to implement $\Y\to
\pi_{\pi/2}\pi_{\phi+\pi/2}(2\pi)_{3\phi+\pi/2}\pi_{\phi+\pi/2}$ as well as
$\bar \X\equiv \X_\pi$, while in the second part of the sequence we used the
same decomposition but backwards, e.g.,
$\X\to\pi_\phi(2\pi)_{3\phi}\pi_\phi\pi_0$.  Here
$\phi=-\cos^{-1}(-1/4)\approx104.5^\circ$.  With no amplitude mismatch, thus
constructed decoupling sequence has order $K=1$ with 1st-order pulses, and an
almost-2nd order ``$1$*'' with 2nd-order pulses, with the norm of the
2nd-order term reduced by three orders of magnitude compared to
1st-order pulses.  These should be compared with $K=1$ and $K=3$ respectively for the
regular 8-pulse sequence [Tab.~\ref{tab:decoup-order}].

It is seen from Fig.~\ref{fig:8_bb1}(a) that regular 1st- and 2nd-order pulses
$S_1$ and $Q_1$ 
perform in the BB$_1$-corrected composite sequence almost as well as the
self-refocusing pulses with
additional amplitude protection (1st-order $L_1$ and 2nd-order $L_3$).  In
addition, the 2nd-order pulses with amplitude protection ($L_3$) work almost
as well in the regular 8-pulse sequence.  All of these allow to achieve the
infidelity level of $10^{-4}$ at amplitude mismatch of $3\%$ or higher (up to
$6\%$ with pulses $L_3$ in the sequence \textbf{8}BB$_1$).  With the regular
8-pulse sequence, the region of high fidelity shrinks substantially for pulses
$Q_1$ and $L_1$, and it all
but disappears for the pulse S$_1$ [the coefficient $\alpha$ for the pulse
$L_1$ happens to be a few times smaller than that for $S_1$].

The results of analogous calculation for a particular 4-qubit XXZ chain with
on-site chemical potential shifts $\Delta^z_i$ over time interval $t=128\tau$
(exactly the same parameters as in Figs.~\ref{fig:samples}, \ref{fig:size})
are presented in Fig.~\ref{fig:8_bb1}(b).  However, it turns out that the
BB$_1$ composite pulses lose accuracy when used with XXZ chain.  Even in the
absence of amplitude errors, there are linear errors in $\tau J^\perp$ [the
zeroth order average Hamiltonian is non-zero].  Thus, we only present the
results for the regular 8-pulse sequence.  With this sequence, the decoupling
orders for 1st- and 2nd-order pulses are $K=1$ and $K=2$ respectively
[Tab.~\ref{tab:decoup-order}].  Compared with Ising-only couplings, with the
particular parameters chosen, this increases the infidelity by some five
orders of magnitude with pulses Q$_1$ [Fig.~\ref{fig:samples}(c)], and by some
two orders of magnitude for pulses $S_1$ [Fig.~\ref{fig:samples}(b)].  As for
the Ising chain, the effect of the amplitude errors is weaker with
specially-designed pulses $L_1$ and $L_3$ such that the 1st-order coefficient
$\tilde\upsilon(f)$ scales as a higher power of $f$.  With the pulse $L_1$,
the 2nd-order coefficient $\alpha$ is non-zero but small; it is seen from
Fig.~\ref{fig:8_bb1}(b) that its effect is to introduce a linear term
$1-F\propto f$ which tends to skew the infidelity minimum away from $f=0$.  We
should also note that with the Gaussian pulses G$_{010}$ (not shown), even the
on-resonance infidelity is out of range of the plots Fig.~\ref{fig:8_bb1}. 

\section{Conclusions}
We presented a comprehensive study targeting pulse and sequence design
and analysis based  on a consistent high-order average Hamiltonian
expansion.  The numerical technique for expanding the evolution
operator was originally introduced by us in
Ref.~\onlinecite{sengupta-pryadko-ref-2005}, and a complimentary
analytical technique was developed for $\pi$-pulses by one of the
authors in Ref.~\onlinecite{pryadko-quiroz-2007}.  

The overall approach is to start with a closed system described by a
finite-dimensional Hamiltonian $H_S$ and design a sequence of shaped
pulses such that the evolution operator would be accurate to a given
order $K$ in powers of $H_S$.  The key to this approach are the
NMR-style 1st- and 2nd-order self-refocusing one-dimensional pulses
constructed for a single-qubit chemical-shift
Hamiltonian~(\ref{eq:chemical-shift}).  In this work we designed a
number of such shapes for different rotation angles $\phi_0$, and
presented a careful analytical analysis of the first two leading
orders of the average Hamiltonian theory for driven qubit evolution
with the most general system Hamiltonian $H_S$.  While any symmetric
one-dimensional pulse shape is characterized by only three parameters,
two of these can be set to zero by pulse shaping.  The remaining
parameter is also non-zero for an ideal ``hard'' $\delta$-function
pulse.  This leads to an important conclusion that the constructed
pulses can be used as drop-in replacement for hard pulses; with proper
pulse placement the results should be identical to first two orders.
The structure of errors appearing in higher orders of the evolution
operator can be understood by analyzing the numerical time-dependent
perturbation series for the evolution operator of a closed system.  

An important advantage of this approach is that the expansion order offers a
natural classification of the error operators.  As a result, ({\bf i}) the
convergence regions have regular shapes as a function of parameters [see
Figs.~\ref{fig:bb1-CLJ}(c,d) and \ref{fig:bb1-Wp}(b,c)].  Furthermore, with
local two- (or few-)qubit couplings dominant, ({\bf ii}) the error operators
can be placed on connected clusters of up to $k+1$ qubits for terms of order
$k$, which allows one to understand their structure in terms, e.g., the direct
products of up to $(k+1)$ Pauli matrices.  Once their structure understood,
the convergence can be readily improved by suppressing the error operators, as
in our analysis of pulse-amplitude errors.  We emphasize, that such an
analysis can be performed even for very large qubit systems.  Thus, ({\bf
  iii}) this approach is characterized by scalability with the system size, as
we illustrated by analyzing decoupling infidelity with the system size
[Fig.~\ref{fig:size}].  Although in this work we concentrated on the dynamics
of closed systems, another important advantage is that ({\bf iv}) the
high-order control sequences result in lower decoherence in the presence of
slow environmental modes\cite{pryadko-sengupta-kinetics-2006,pryadko-quiroz-2007}.

Most obvious application of highly-optimized shaped pulses of the sort
presented in this work is in solid-state quantum computation, where
the bandwidth available for quantum gates is typically limited. 
Our techniques based on analytical and numerical high-order average
Hamiltonian theory offers a systematic scalable approach for
constructing gates for such multi-qubit systems, without need of
solving their full dynamics.  However, even if the bandwidth does not
appear to be at premium, simple pulse shaping (e.g., using 1st-order
pulses) can still offer a substantial improvement of control accuracy.

\section{Acknowledgments.} This research was supported in part by the NSF
grant No.\ 0622242 (LP). LANL is supported by US DOE under Contract No. W-7405-ENG-36. 
NHMFL is supported by the DOE, the NSF, and the state of Florida.

\appendix
\section{Average fidelity}
Here we discuss the calculation of the fidelity averaged over the initial
state, in the case of unitary evolution with known evolution matrix $U$, while
the desired evolution matrix is $U_0$.  Let
us write the density matrix of the initial pure state as
$\rho_0=\psi\psi^\dagger$, where $\psi$ is an $N$-component complex vector.
Then the actual density matrix is $\rho=U\psi\psi^\dagger U^\dagger$,
while the desired density matrix is $\rho_\mathrm{ideal}=U_0\psi\psi^\dagger
U_0^\dagger$. 
The fidelity with the given initial state
\begin{eqnarray}
  \label{eq:fidelity-given}
  \nonumber
  F_\psi&\equiv& \tr (\rho_\mathrm{ideal} \rho)=\tr(U_0\psi\psi^\dagger
  U_0^\dagger U\psi\psi^\dagger
  U^\dagger)\\&=&\sum_{ijkl}\psi_i\psi_j^*(U_0^\dagger U)_{ik}\psi_k\psi^*_l
  (U_0^\dagger U)^*_{jl}.
\end{eqnarray}
The only condition on the components $\psi_i$ of the wavefunction is the
normalization, $1=\psi^\dagger\psi=\sum_{i}|\psi_i|^2$.  Generally, this means
that the average of the product in Eq.~(\ref{eq:fidelity-given}) can only
depend on the identity tensor $\delta_{ij}$.  By symmetry, $\langle
\psi_i\psi_j^*\psi_k\psi^*_l\rangle
=A(\delta_{ij}\delta_{kl}+\delta_{il}\delta_{kj})$, where the unknown
coefficient $A$ can be computed from the normalization by tracing over $i=j$,
$k=l$.  We obtain $1=A(N^2+N)$, so that the average fidelity 
\begin{equation}
  \label{eq:average-fidelity}
  F={N+|\tr V|^2\over N+N^2},\quad 1-F={N^2-|\tr V|^2\over N+N^2},
\end{equation}
where $V\equiv U_0^\dagger U$.  Numerically, with $V$ close to identity
matrix, the loss of precision can be avoided by expressing the infidelity
$1-F$ in terms of the modified mismatch,
\begin{equation}
  \label{eq:mismatch}
  \delta^2=\tr(\openone-\tilde V)^\dagger (\openone-\tilde V),\quad 
  \tilde V=V {|\tr V|\over \tr V}.
\end{equation}
Namely, since $\delta^2=2N-2|\tr V|$, the average infidelity can be written as 
\begin{equation}
  \label{eq:average-infidelity-delta}
  1-F={\delta^2 (4N-\delta^2)\over 4 (N+N^2)}.
\end{equation}


\end{document}